\documentclass[prl,amsmath,amssymb,aps,10pt,superscriptaddress,longbibliography,twocolumn,floatfix]{revtex4-2}
\usepackage{amsmath, amsthm, amssymb, amsfonts}
\usepackage{graphicx} 
\usepackage[colorlinks=true, linkcolor=blue, urlcolor=blue, citecolor=blue, anchorcolor = blue]{hyperref} 
\usepackage{color,xcolor}
\usepackage{dcolumn}
\newcolumntype{d}[1]{D{.}{.}{#1}}
\usepackage[utf8]{inputenc}
\usepackage{pdfpages}
\usepackage{pgffor} 
\usepackage{xr} 
\makeatletter
\AtBeginDocument{\let\LS@rot\@undefined}
\makeatother

\def\supplementfilename{supplemental_material_plain}

\pdfximage{\supplementfilename.pdf}
\def\numbersupplementpages{\the\pdflastximagepages}

\newif\ifarXiv
\arXivtrue 

\usepackage{soul} 
\usepackage[normalem]{ulem}

\definecolor{orange}{rgb}{0.99,0.5,0.31}

\newlength{\lengthofminus}
\newcommand{\dl}{\hspace{\lengthofminus}} 

\newlength{\lengthofone}
\begin{document}

\title{Ultra precise determination of Cs($nS_{1/2}$) and Cs($nD_J$) quantum defects for sensing and computing: Evaluation of core contributions}

\date{\today}

\author{Pinrui Shen}
\affiliation{ Quantum Valley Ideas Laboratories, 485 Wes Graham Way, Waterloo, Ontario N2L 6R1, Canada}
\author{Donald Booth}
\affiliation{ Quantum Valley Ideas Laboratories, 485 Wes Graham Way, Waterloo, Ontario N2L 6R1, Canada}
\author{Chang Liu}
\affiliation{ Quantum Valley Ideas Laboratories, 485 Wes Graham Way, Waterloo, Ontario N2L 6R1, Canada}
\author{Scott Beattie}
\affiliation{ 
National Research Council Canada, Ottawa, Ontario K1A 0R6, Canada
}
\author{Claude Marceau}
\affiliation{ 
National Research Council Canada, Ottawa, Ontario K1A 0R6, Canada
}

\author{James P. Shaffer}
\email{e-mail: jshaffer@qvil.ca}
\affiliation{ Quantum Valley Ideas Laboratories, 485 Wes Graham Way, Waterloo, Ontario N2L 6R1, Canada}

\author{\\ Mariusz Pawlak}
\affiliation{ 
Faculty of Chemistry, Nicolaus Copernicus University in Toru\'n, Gagarina~7, 87-100~Toru\'n, Poland
}

\author{H. R. Sadeghpour}
\email{e-mail: hrs@cfa.harvard.edu}
\affiliation{ ITAMP, Center for Astrophysics $|$ Harvard \& Smithsonian, 60 Garden Street., Cambridge, Massachusetts 02138, USA
}

\begin{abstract}

We make absolute frequency measurements of Cs Rydberg transitions, $\vert 6S_{1/2}, F=3 \rangle \rightarrow \vert nS_{1/2}$~$(n=23\rm{-}90)\rangle$ and $\vert nD_{3/2,5/2}$~$(n=21\rm{-}90)\rangle$, with an accuracy of less than $ 72\,\rm kHz$. The quantum defect parameters for the measured Rydberg series are the most precise obtained to date. The quantum defect series is terminated at $\delta_4$, showing that prior fits requiring higher order quantum defects reflect uncertainties in the observations. The precision of the measured quantum defects allow for the calculation of Rydberg electric-dipole transitions and fine-structure intervals extrapolated from high principal quantum numbers, to rival that of sophisticated many-body relativistic calculations carried out at low Rydberg principal quantum numbers. We quantitatively predict the contributions to the quantum defect parameters from core polarization and core penetration of Cs inner shell electrons. A new value for the ionization energy, consistent across the $ nS_{1/2}$ and $ nD_{3/2,5/2}$ Rydberg series, is reported at $31406.467 751 48 (14)~\rm{cm}^{-1}$.  

\end{abstract}
\maketitle


Neutral-atom-based quantum simulators and processors are the leading architectures for high fidelity quantum operations while atoms are being used extensively as sensors because of their stability and uniformity. Rydberg atoms are particularly suitable for quantum simulations and metrology, because their properties can be tuned through state selection and the use of electromagnetic fields. Precise knowledge of atomic structure enables engineering of atomic properties.
Similarly for basic research, precise measurements of atomic energy levels provide stringent tests of atomic structure calculations, including quantum defect theories, and are critical for the determination of a number of fundamental constants in physics \cite{Safronova2018,Myers2019,Pritchard2009}. Cesium atoms have been used for searches of the electric-dipole moment of the electron~\cite{Weisskopf1968,Murthy1989,Bernreuther1991,Chin2001}, among other fundamental quantities. Rapid  developments in quantum measurement technologies, such as optical clocks~\cite{Wilpers2002,Pan2020,Ludlow2015,Sharma2022,Bothwell2022}, neutral-atom-based quantum computing \cite{Graham2019,Saffman2010,Bluvstein2024,Henriet2020,Pause2024} and Rydberg-atom-based electric field sensors~\cite{Sedlacek2012,Fan2015,Fancher2021} are adding impetus to accurately determine energy levels as well as derived quantities such as electric field polarizabilities and reduced electric-dipole matrix elements. The aim of this work is to improve the accuracy of Cs atomic energy level measurements for advanced applications in quantum sensing.

We present absolute frequency measurements for the Cs transitions $\vert 6S_{1/2}, F=3 \rangle\rightarrow \vert nS_{1/2}$~$(n=23\rm{-}90)\rangle$ and $\vert nD_{3/2,5/2}$~$(n=21\rm{-}90)\rangle$ with an accuracy of $<~72\,\rm kHz$. By fitting the absolute-frequency measurements to the modified Ritz formula, we determine the most accurate quantum defect parameters  of the $nS_{1/2}$, $nD_{3/2}$, and $nD_{5/2}$ series to date, and obtain a new ionization energy, $31406.467 751 48 (14)~\rm{cm}^{-1}$. Because of the accuracy of the underlying term energies, the highest order fitted quantum defect is $\delta_4$, illustrating that the higher terms, often found in the literature, reflect uncertainties in the measurements. We extract the fine-structure intervals for the $nD_J$ series. The precision of our calculations of electric-dipole transition matrix elements is comparable with that of sophisticated many-body relativistic calculations \cite{Safronova2016} at low $n$, where measurements were not carried-out. We calculate the contributions to the quantum defect parameters from core polarization and core penetration. Both effects are found to be in agreement with high level atomic structure calculations.

The experimental cycle begins by collecting Cs atoms in a magneto-optical trap (MOT). The MOT atoms are cooled to $< 10~\mu\rm K$ using polarization gradient cooling~\cite{Dalibard1989,Ji2014}. After the cooling stage, the atoms are optically pumped to the $|6S_{1/2}, F=3\rangle$ ground state. Subsequently, the trapping magnetic field and lasers are switched off to create a field-free environment for the atoms. During the entire interrogation period, the residual magnetic field is reduced to $< 2~\rm mG$ and the stray electric field is reduced to $< 1$~mV/cm, using external field compensation \cite{SM}.

A two-step laser excitation at $\sim 852~\rm nm$ and $\sim 509~\rm nm$ creates cold Cs Rydberg atoms. The lasers are both locked to a frequency comb that is stabilized to a GPS steered rubidium clock, resulting in laser linewidths of less than 200~Hz and a fractional frequency stability of $1.2\times10^{-12}$ for a $10~\rm s$ averaging time. The $\sim 852~\rm nm$ laser is red-detuned 90~MHz from the $|6P_{3/2}$, $F'=2\rangle$ excited state. The $\sim 509~\rm nm$ laser is 90~MHz blue-detuned from the resonance between the $|6P_{3/2}$, $F'=2\rangle$ state and the target $nD_J$ or $nS_{1/2}$ Rydberg state. The two lasers are crossed in a perpendicular configuration, both with parallel linear, vertical polarizations. The Gaussian beam waists for the first- and second-step lasers are $w_0=3~\rm{mm}$ and $w_0=400~\mu\rm{m}$, respectively. The lasers are pulsed to excite the Rydberg atoms. The duration of the excitation laser pulses is $30~\mu$s. The pulses overlap in time.

Following the laser excitation pulses, a 2~$\mu$s high voltage square pulse is applied to field ionization plates to ionize the Rydberg atoms and project the ions onto a~multi-channel plate (MCP) detector. The average number of detected ions generated per laser shot is minimized to $<2$ to reduce Coulomb repulsion. The ion pulses are recorded on a multichannel analyzer.

To obtain the Rydberg state spectrum, the frequency of the $\sim 852~\rm nm$ laser is fixed and the frequency of the $\sim 509~\rm nm$ laser is scanned. The laser excitation and ion detection sequence is repeated 40 times for each cold atomic cloud. Twenty clouds are used for each $\sim 509~\rm nm$ laser frequency, resulting in 800 excitation-detection cycles per laser frequency step.  The spectroscopy measurements are performed for the $nD_{5/2}$~($n=21$--90), $nD_{3/2}$~($n=21$--90), and $nS_{1/2}$~($n=23$--90) series.

For the $nD_{3/2}$ and $nD_{5/2}$ series, the measured spectrum is fit to a~hyperfine-structure-weighted multiline Gaussian function \cite{SM}. For the $nS_{1/2}$ series spectrum, a~single Gaussian function is used since there is only one allowed hyperfine transition, $\vert 6P_{3/2}, F=2 \rangle \rightarrow \vert nS_{1/2}, F'=3 \rangle$, and the dominant broadening mechanism is Doppler broadening, at the $200~\rm{kHz}$ level. The spectral line frequency is the absolute transition frequency between the $|6S_{1/2}, F=3\rangle$ hyperfine state and the center of gravity of the Rydberg energy level.

The transition frequencies are corrected to account for the AC Stark shifts caused by the lasers. The AC Stark shift induced by the $\sim~509~\rm nm$ laser is no greater than $500~\rm{Hz}$ for the $21D_{J}$ state. The $\sim~852~\rm nm$ laser induces a~measured ($36.0\pm4.8)~\rm{kHz}$ blue shift of the ground state, $|6S_{1/2}, F=3\rangle$ \cite{SM}. The 36.0~kHz shift is subtracted from the transition frequency and the uncertainty, $\pm 4.8~\rm kHz$, is taken as a contribution to the uncertainty of the resonance frequency.

For the $nS_{1/2}$ series, the DC Stark effect induces a Rydberg level frequency shift. The maximum shift is calculated to be $<1.5~\mathrm{kHz}$ for the $90S_{1/2}$ state \cite{ARC2017,Meyer2020,Bai2020}. For the $nD_J$ series, the DC Stark effect predominantly splits the energy levels because different $m_J$ states have different Stark shifts. It is difficult to resolve the peak splitting when the stray electric field is small, $< 1~\rm mV/cm$. Thus, the DC Stark effect causes asymmetric line broadening of the $nD_J$ states and is treated as a contribution to the uncertainty of the measured transition frequency. Based on the polarizability of the Cs $nD_{5/2}$ states, the maximum broadening is estimated to be $<49~\mathrm{kHz}$ for the  90$D_{5/2}$ state~\cite{SM,Auzinsh2007}.

The Zeeman shift of the ground state is minimized using microwave spectroscopy \cite{SM,Sassmann2013}. The Zeeman shift was measured to be $<14.3~\rm kHz$. The sensitivity to the magnetic field is similar across all Rydberg states because they have the same Land{\'e} $g$-factor. The Zeeman shift is treated as another source of uncertainty.

\begin{figure}[!ht]
    \centering
    \includegraphics[width=0.42\textwidth]{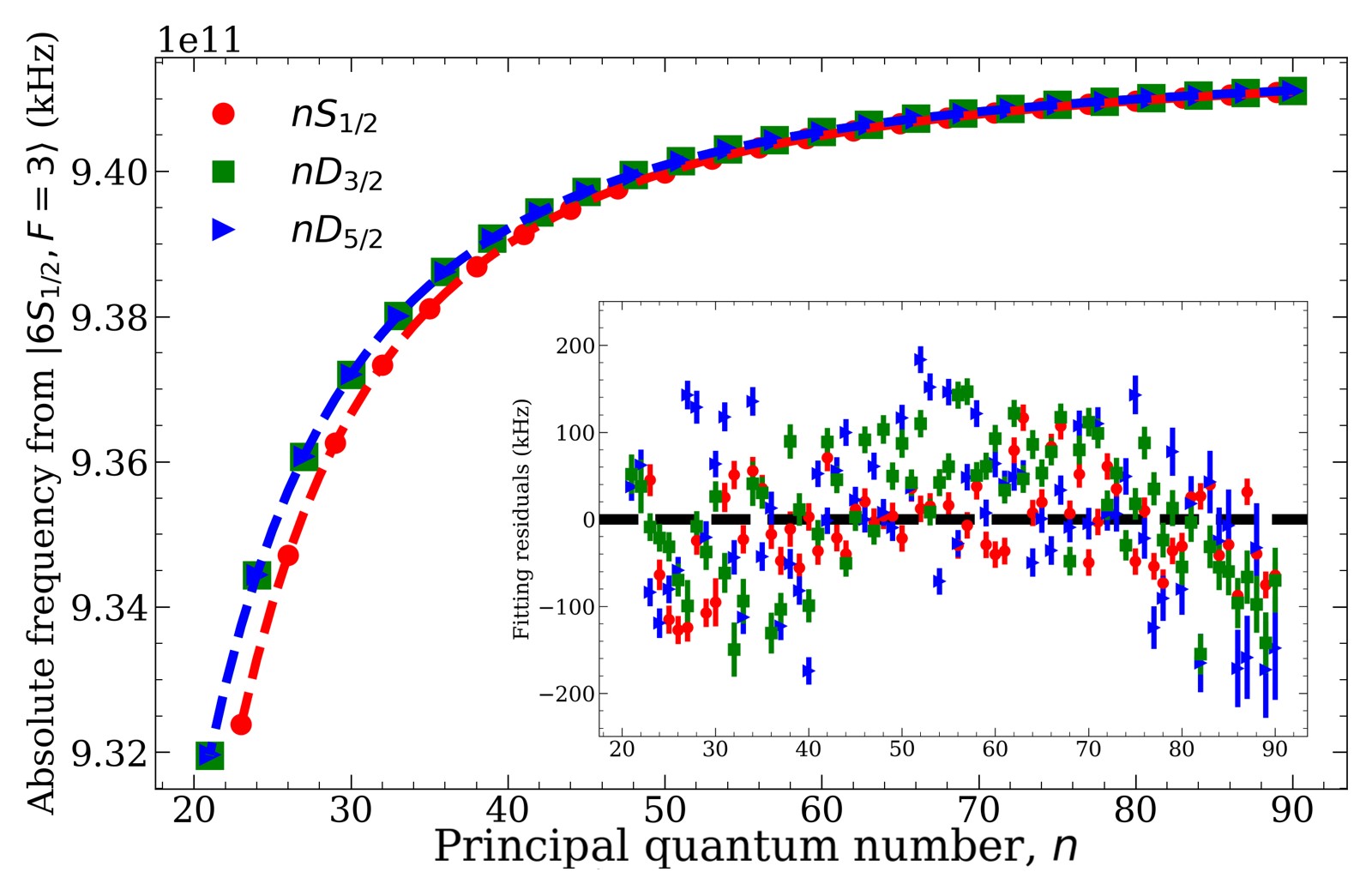}
    \caption{Absolute frequency measurements for transitions from $|6S_{1/2}, F=3\rangle$ to the center of gravity of the $nS_{1/2}$, $nD_{3/2}$, $nD_{5/2}$ Rydberg states. Data are presented with a~step of $\Delta n=3$ to avoid crowding. The dashed lines correspond to the modified Ritz formula. The fitting residuals are shown in the inset. The error bars represent the total uncertainty. The statistical variance of the fitting residuals of the $nD_{5/2}$, $nD_{3/2}$, and $nS_{1/2}$ series are $71.2~\rm kHz$, $59.1~\rm kHz$, and $46.9~\rm kHz$, respectively.}
    \label{fig:absfrequency}
\end{figure}

The shot-to-shot variance dominates the statistical uncertainty. Systematic shifts, most probably associated with residual Stark shifts, can be observed in the plot of the residuals, Fig.~\ref{fig:absfrequency}. The sum of the statistical uncertainty and the systematic uncertainty is no greater than 60~kHz \cite{SM}. The statistical standard deviation of the residuals of the modified Ritz formula fit is $< 72~\rm kHz$ for all series.

\begin{table*}[!ht]
    \centering
        \caption{Quantum defects for the $nD_{5/2}$, $nD_{3/2}$, and $nS_{1/2}$ series obtained by fitting the measured Rydberg energy levels to Eq.~(\ref{Eq:QDequation}). The uncertainties are the standard deviations of the fit. For comparison, the quantum defects used in the ARC python calculation package \cite{ARC2017} are also listed. }
    \scalebox{0.85}{
    \begin{tabular}{c|c|c|c|c|c|c}
    \hline
    \hline
         &$nD_{5/2}$ [this work] & $nD_{5/2}$ \cite{Weber1987} &$nD_{3/2}$ [this work] & $nD_{3/2}$ \cite{Lorenzen1984} &$nS_{1/2}$ [this work] & $nS_{1/2}$ \cite{Weber1987}\\ 
         &($n=21$--90) & ($n=5$--36) & ($n=21$--90) &($n=5,6,8$--32) & ($n=23$--90) & ($n=6$--30) \\    \hline      
         IE $(\rm{cm}^{-1})$& 31406.46775168(26)&31406.46767(3)&31406.46775148(28)&31406.4710(7)  &31406.46775136(21) & 31406.46769(2) \\  \hline\hline
         $\delta_0$   &   2.46631529(14) &   2.46631524(63)&   2.47545840(13)&   2.4754562& 4.04935994(17)&4.04935665(38)\\ 
         $\delta_2$   &   0.013431(65)   &   0.013577      &   0.008339(62) &   0.009320 & 0.238017(18)  &0.2377037\\
         $\delta_4$   &$-$0.3613(61)     &$-$0.37457       &$-$0.3769(95)   &$-$0.43498  & 0.1747(33)    &0.255401\\
         $\delta_6$   &   0.0            &$-$2.1867        &   0.0          &$-$0.76358  & 0.0           &0.00378\\
         $\delta_8$   &   0.0            &$-$1.5532        &   0.0          &$-$18.0061  & 0.0           &0.25486\\
         $\delta_{10}$&   0.0            &$-$56.6739       &   0.0          &    0.0     & 0.0           &0.0\\
         \hline \hline
    \end{tabular}}
    \label{tab:quantdef}
\end{table*}

\begin{table}
    \centering
        \caption{ Numerically computed core polarization contribution, $\delta_{\rm pol}^{(l,J)}$, and core penetration contribution, $\delta_{\rm pen}^{(l,J)}$, to the leading quantum defect term of the $nD_{5/2}$, $nD_{3/2}$, and $nS_{1/2}$ series.}
    \scalebox{0.85}{
    \begin{tabular}{l|l|l|l}
    \hline
    \hline
         \multicolumn{1}{c|}{series}  & \multicolumn{1}{c|}{$\delta_0^{(l,J)}$}  & \multicolumn{1}{c|}{$\delta_{\rm pol}^{(l,J)}$} & \multicolumn{1}{c}{$\delta_{\rm pen}^{(l,J)}$}  \\  \hline
         $nD_{5/2}$ &  2.46631529(14)  & 1.0482(4) & 1.4197(6)\\ 
         $ nD_{3/2}$ & 2.47545840(13)   & 1.0625(7) & 1.4094(5)\\
         $nS_{1/2}$ & 4.04935994(17)  & 0.1320(5) & 3.9111(4)\\
         \hline \hline
    \end{tabular}}
    \label{tab:QDcontributions}
\end{table}


The observed term energies of a single Rydberg series can be described by the modified Ritz formula:
\begin{equation}\label{Eq:QDequation}
E_{n,l,J} =  E_I - \frac{R}{[n-\delta^{(l,J)}(n)]^2},
\end{equation} 
where $E_I$ is the ionization energy and $R$ is the reduced mass Rydberg constant, equal to $R_{\infty}/(1+m_{\rm e}/m_{\rm core})$. Using the current mass of the Cs atom \cite{CIAAW}, we obtained $R_{^{133}\rm Cs} = $ 109736.8627304~cm$^{-1}$. The $(n,l,J)$-dependent quantum defects are
\begin{equation}\label{eq:QDexpan}
\delta^{(l,J)}(n) = \delta^{(l,J)}_0 + \sum_{k=1}^{\infty} \frac{\delta^{(l,J)}_{{2k}}}{[n-\delta^{(l,J)}_0]^{2k}}.
\end{equation}
We constrain the quantum defects to a three-term expansion, $\delta^{(l,J)}(n)=\delta^{(l,J)}_0 + \frac{\delta^{(l,J)}_{{2}}}{[n-\delta^{(l,J)}_0]^{2}}+ \frac{\delta^{(l,J)}_{{4}}}{[n-\delta^{(l,J)}_0]^{4}}$, and fit the measured energy levels to Eq.~(\ref{Eq:QDequation}), which minimizes the correlations between the ionization energy and the quantum defects present in the fit. The Rydberg state energy is obtained by subtracting the frequency splitting, $5.170855370625~\rm GHz$, between the center of gravity of the $6S_{1/2}$ state and the $|6S_{1/2}, F=3\rangle$ hyperfine state~\cite{steckCsdata} from the measured absolute transition frequencies listed in Table~S4~\cite{SM}, i.e., the energies are referenced to the ground state center of gravity to perform the fitting.

The ionization energies are reported in Table~\ref{tab:quantdef}. The extracted ionization energies for the $nD_{5/2}$, $nD_{3/2}$, and $nS_{1/2}$ states agree within their uncertainties. The uncertainty-weighted average ionization energy is $31406.467 751 48 (14)~\rm{cm}^{-1}$. 
The result agrees with the ionization energy reported in Ref.~\cite{Weber1987}, but is $600~\rm kHz$ higher than the most recent value presented in Ref.~\cite{Deiglmayr2016}, $31406.467 732 5 (14)~\rm{cm}^{-1}$. In Ref.~\cite{Deiglmayr2016}, Cs samples were prepared in an optical dipole trap that cannot be completely shut-off, leading to a larger AC Stark effect. The ionization energy reported in Ref.~\cite{Deiglmayr2016} was extracted using one Rydberg series, $\vert nP_{J}$~$(n=27\rm{-}74)\rangle$ where the $n$ step size was 3. In this work, the Cs samples are excited in a field-free environment and the ionization energy is extracted from all of the levels in the three Rydberg series $\vert nS_{1/2}$~$(n=23\rm{-}90)\rangle$  and $\vert nD_{J}$~$(n=21\rm{-}90)\rangle$. Moreover, the uncertainty of our ionization energy determined by the fit to the modified Ritz formula is further reduced by one order of magnitude.

The non-linear least squares fit to measured energies are shown in Fig.~\ref{fig:absfrequency} with residues in the inset. The systematic downward trend for the higher principal quantum numbers of the $nD_J$ series is likely caused by DC Stark shifts. The quantum defects for all three series are presented in Table~\ref{tab:quantdef}. Our results show that the $\delta_{2k}^{(l,J)}$ parameters for $k>2$ for all three series are insignificant within the measurement uncertainties. Several studies have reported these terms for the same states as non-zero at lower levels of experimental precision \cite{Weber1987, Lorenzen1984}. Since the precision of our measurements is two orders of magnitude better than that of the prior work, it is likely that these terms have no physical significance and were fitting parameters that accommodated experimental uncertainties \cite{SM}.

\begin{figure}[!ht]
    \centering
    \includegraphics[width=0.42\textwidth]{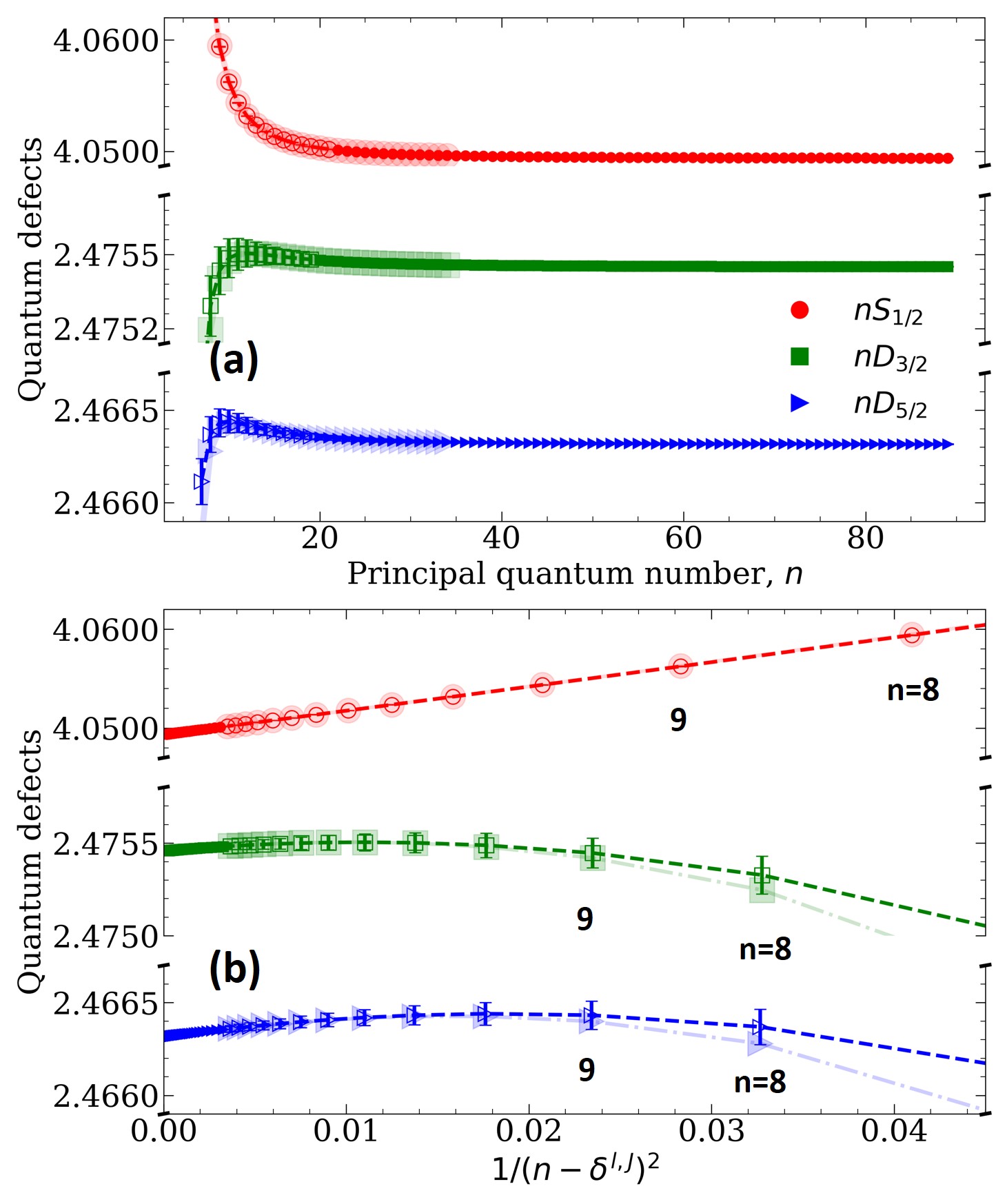}
    \caption{Experimental quantum defects, $\delta^{(l,J)}$, versus $n$ (a) and $1/[n-\delta^{(l,J)}]^2$ (b), for all measured term series in Cs. Solid markers are the quantum defects measured in this work, while open markers are extrapolated from the experimental measurements. The light color markers represent the values calculated based on previous publications, $nS_{1/2}$ \cite{Weber1987}, $nD_{3/2}$~\cite{Lorenzen1984}, and $nD_{5/2}$ \cite{Weber1987}. The error bars are acquired by propagating the fitting uncertainty of the quantum defects. }
    \label{fig:QDvsN}
\end{figure}

The quantum defect parameters obtained from the measurements in Fig.~\ref{fig:QDvsN}(a) agree well within the uncertainties, even for lower-lying states ($8<n<21$), where no measurements were taken. The plots show further evidence that the quantum defect expansion greater than third order is not required, since our predictions for the $nS_{1/2}$, $nD_{3/2}$, and $nD_{5/2}$ energy levels match the available experimental data at low $n$. Notably, the trend for both $nD_J$ series is non-monotonic while that of the $nS_{1/2}$ series always decreases as a function of $n$. The non-monotonic behavior of the $nD_J$ series had been reported in Refs.~\cite{Lorenzen1983, Lorenzen1984, Weber1987}. The behavior can be viewed more clearly by plotting the quantum defects as a function of $1/[n-\delta^{(l,J)}]^2$, shown as Fig.~\ref{fig:QDvsN}(b). The difference between $nS_{1/2}$ and $nD_J$ is the result of a competition between core polarization and core penetration. Core polarization reduces the quantum defect, more so as the electron effective radius is reduced, i.e., low $n$. As $n$ increases, the core polarization effect decreases faster than the core penetration effect, which leads to a quantum defect dominated by core penetration.

Weber {\it et al.}~\cite{Weber1987} investigated the contribution from core polarization for the $nD_J$ series of Cs and concluded that it is too small to account for the non-monotonic behavior. We quantitatively studied the behavior by numerically calculating the contributions to the quantum defect's leading term from core polarization and core penetration. Using the model potential in Ref.~\cite{Marinescu1994} in conjunction with the spin--orbit coupling effect, we calculated the expectation values of the energy due to core polarization and core penetration~\cite{SM}. 
The numerically determined contributions to $\delta^{(l,J)}_0$ are presented in Table~\ref{tab:QDcontributions}. For the $nD_J$ series, the contribution from core polarization is over 40\%, much larger than the estimate in Ref.~\cite{Weber1987}, possibly because the second-order dipole polarization corrections were not included \cite{Drake1991}. If core polarization is neglected, the $nD_J$ quantum defects become larger as $n$ decreases, similar to $nS_{1/2}$. For the $nS_{1/2}$ series, the contribution from core polarization is less than 2.5\%.

\begin{table}[!b]
    \centering
        \caption{Fine-structure interval parameters determined for $nD_J$ series both experimentally (Exp) and numerically (Num). The quoted uncertainties are the standard deviations of the fit.}
\scalebox{0.85}{
    \begin{tabular}{@{}l@{}|l|l|l@{}}    
    \hline \hline
        $nD_{3/2} \rightarrow nD_{5/2}$ & \multicolumn{1}{c|}{$\xi^{(l)}_1$ (kHz) } & \multicolumn{1}{c|}{$\xi^{(l)}_2$ (kHz)} & \multicolumn{1}{c}{$\xi^{(l)}_3$ (kHz)} \\ \hline 
        Exp [this work]              & 6.01624(22)$\times10^{10}$ & $-$3.38(24)$\times10^{10}$ & 8(6)$\times10^{11}$ \\  
        Exp \cite{Goy_PRA_1982} & 6.02183(60)$\times10^{10}$ & $-$5.8(8)$\times10^{10}$   & \multicolumn{1}{c}{$0$} \\         
        Num [this work]     & 6.547$\times10^{10}$    & \dl 5.5 $\times10^{10}$ & 7.7$\times10^{11}$ \\         
        \hline\hline
    \end{tabular}}
    \label{tab:finespara}
\end{table}


We studied the fine-structure splitting of the $nD_J$ series. The fine-structure splitting, $\Delta_{\rm fs}$, can be expanded in terms of a power series in the effective principal quantum number, [$n-\tilde{\delta}^{(l,J=l\pm1/2)}$], \cite{Goy_PRA_1982}: 
\begin{equation}\label{eq:finestrucsep}
\Delta_{\rm fs} = \sum_{q=1}^{\infty} \frac{\xi^{(l)}_q}{[n-\tilde{\delta}^{(l,J=l\pm1/2)}(n)]^{2q+1}}.
\end{equation}
The weighted average quantum defect, $\tilde{\delta}^{(l,J=l\pm1/2)}(n)$,  has the form
\begin{equation}\label{eq:averagedelta}
\tilde{\delta}^{(l,J=l\pm1/2)}(n) = w^{(l,J<l)} {\delta}^{(l,J<l)}(n) + w^{(l,J>l)} {\delta}^{(l,J>l)}(n).
\end{equation}
The weights for the fine-structure splitting, $w^{(l,J)}$, are equal to
$1-\frac{|\langle {\textbf{L}}\cdot{\textbf{S}} \rangle^{(l,J)}|}{\hbar^2(l+1/2)}$,  
where $\langle {\textbf{L}}\cdot{\textbf{S}} \rangle^{(l,J)} = \frac{\hbar^2}{2}[J(J+1)-l(l+1)-\frac{3}{4}]$.
For the $nD_J$ series, $w^{(2,3/2)}=0.4$ and $w^{(2,5/2)}=0.6$.

\begin{figure}[!t] 
    \centering
    \includegraphics[width=0.42\textwidth]{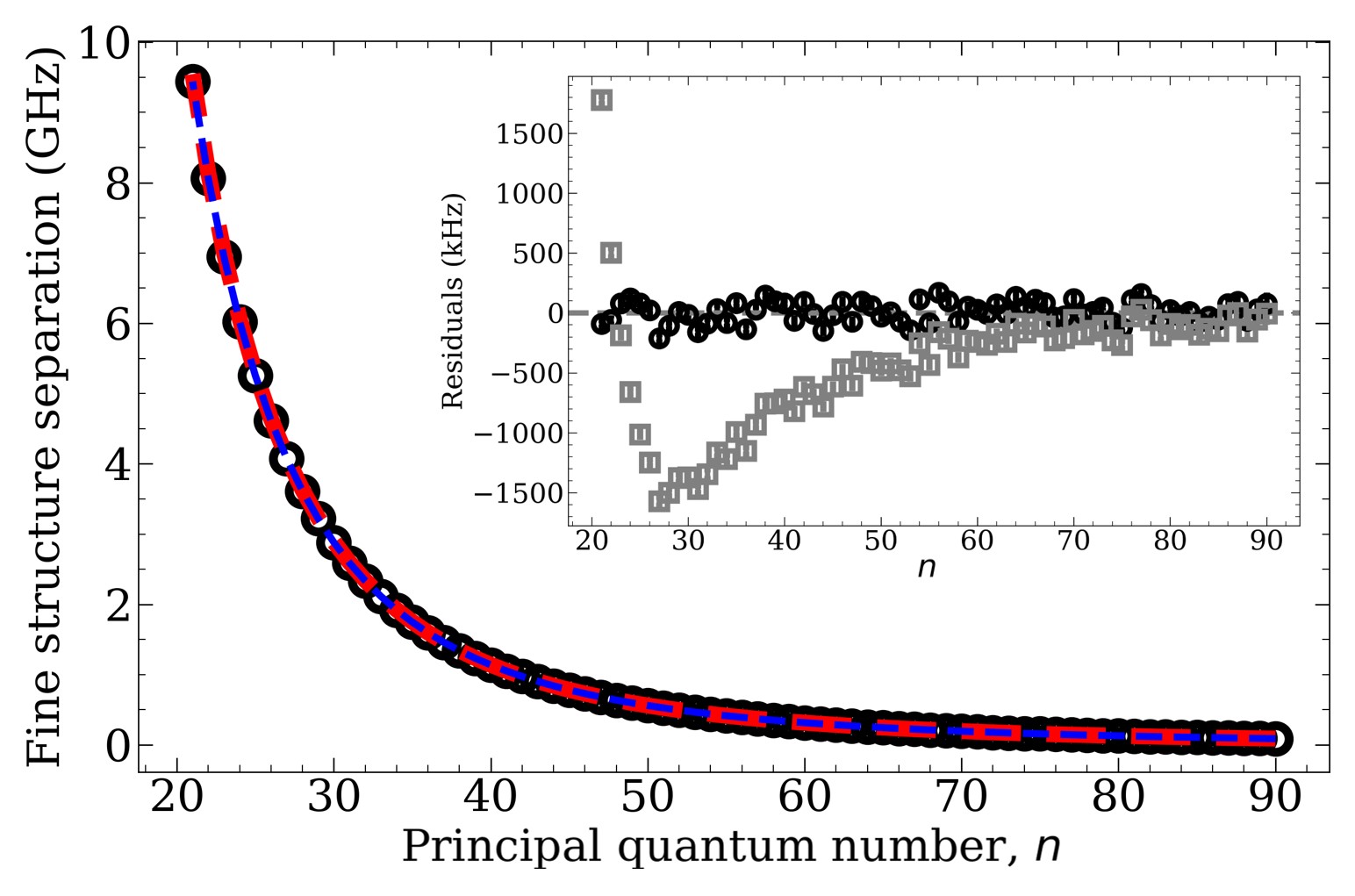}
    \caption{The fine-structure splitting for $nD_J$. The red dashed line shows the prediction using Eq.~(\ref{eq:finestrucsep}). The fitting residuals are given as the black circles in the inset. The fine-structure interval based on the fitting coefficients reported in Ref.~\cite{Goy_PRA_1982} is plotted as the blue dashed line for comparison. The corresponding residuals in data in Ref.~\cite{Goy_PRA_1982}, are shown as gray squares in the inset. }
    \label{fig:finestructuresep}
\end{figure}

Using the quantum defect values in Table~\ref{tab:quantdef} and Eq.~(\ref{eq:averagedelta}), 
the measured fine-structure intervals were fit to Eq.~(\ref{eq:finestrucsep}) to extract $\xi^{(l)}_q$ up to $q=3$. The values of $\xi^{(l)}_q$ were reported in Ref.~\cite{Goy_PRA_1982} based on a single fine-structure interval measurement. The parameters obtained from the measurements presented here are given in Table~\ref{tab:finespara}. The quality of the fit is evident in Fig.~\ref{fig:finestructuresep}; the extracted parameters better describe the fine-structure splittings, $\Delta_{\rm fs}$, at low $n$. The fitting residuals are randomly distributed around zero as shown in the inset of Fig.~\ref{fig:finestructuresep}.

The use of Eq.~(\ref{eq:finestrucsep}) to describe $\Delta_{\rm fs}$ had been proposed and used in previous work \cite{Goy_PRA_1982}. The physical meaning of $\xi^{(l)}_q$, have not been discussed to our knowledge. Here, we justify the expansion~(\ref{eq:finestrucsep}) by analytically deriving the spin--orbit coupling constant with hydrogenic wave functions, $ \langle \frac{1}{r}\frac{{\rm d} V(r)}{{\rm d} r} {\textbf{L}}\cdot{\textbf{S}}  \rangle$, for the atomic potential in the form of the sum of the Coulomb and polarization potentials, see Ref.~\cite{SM}. Moreover, using the effective non-local model potential from Ref.~\cite{Marinescu1994} and the precise quantum defects, accurate wave functions and spin--orbit energies for the $nD_{5/2}$ and $nD_{3/2}$ series are determined. The numerical predictions $\Delta_{\rm fs}$ are obtained by taking the differences between the expectation values of the spin--orbit coupling operator of the two series~\cite{SM}. They are fit to Eq.~(\ref{eq:finestrucsep}) to extract $\xi^{(l)}_q$, as shown in Table~\ref{tab:finespara}. 
The difference in the sign of $\xi^{(l)}_2$ may result from the  model potential adopted for the numerical calculations. Importantly, the values of the leading coefficient for both the experimental and numerical approaches, $\xi^{(l)}_1$, are of similar order of magnitude.
We note that the weighted average quantum defect, $\tilde{\delta}^{(l,J=l\pm1/2)}(n)$, should be weighted by the total angular momenta of the $nD_{5/2}$ and $nD_{3/2}$ series, rather than being weighted equally as was done in Ref.~\cite{Goy_PRA_1982}.

Using our quantum defects and those of Ref.~\cite{Deiglmayr2016} for the $P$ states, we computed reduced electric-dipole matrix elements for $nP_J \leftrightarrow n^{\prime} D_{J^\prime}$ transitions of Cs. The calculated matrix elements for the low-lying Rydberg states agree within theoretical uncertainties, $< 3\%$, with high-precision, relativistic, all-order many-body calculations of Safronova {\it et al.}~\cite{Safronova2016}, which are computationally sophisticated and resource intensive to perform. With accurate quantum defects, garnered from high $n$ states, accurate wave functions and matrix elements can be obtained for all $n$, notably for low $n$ where the spectroscopy is difficult to carry-out because of the energy separations between different states. The full comparison table of the matrix elements for the $nP_J \leftrightarrow n^{\prime} D_{J^\prime}$ transitions is presented in the supplemental material, Table~S3 \cite{SM}.


In summary, we presented high precision, absolute frequency measurements of Cs energy levels from the $\vert 6S_{1/2}, F=3 \rangle$ hyperfine ground state to $\vert nS_{1/2}$~$(n=23-90)\rangle$, $\vert nD_{3/2}$~$(n=21-90)\rangle$, and $\vert nD_{5/2}$~$(n=21-90)\rangle$ Rydberg states with an accuracy of $< 72\,\rm kHz$. We investigated the validity of the modified Ritz expansion and conclude that a modified Ritz expansion with only even quantum defect terms can accurately describe the energy levels of the Cs $nS_{1/2}$ and $nD_J$ states. By fitting the absolute-frequency measurements to the modified Ritz formula, Eq.~(\ref{Eq:QDequation}), we determine the quantum defects of the  Cs $nS_{1/2}$, $nD_{3/2}$, and $nD_{5/2}$ series and the ionization energy. The ionization energy was found to be $31406.467 751 48 (14)~\rm{cm}^{-1}$. At the current level of precision, we find the quantum defect expansion can be terminated at the third order. We separated the contributions from core polarization and core penetration. For the $nD_J$ series, the contribution from core polarization is about 40\%, while for the $nS_{1/2}$ series, the contribution is less than 2.5\%. Using improved Cs Rydberg wave functions derived from the quantum defects, we recalculated reduced electric-dipole matrix elements for the $nP_J \leftrightarrow n^{\prime} D_{J^\prime}$ transitions. The obtained matrix elements are found to be in agreement with high level, many-body relativistic calculations at low $n$, where our measurements were extrapolated.

Some aspects of prior work that were not accurate have been corrected, such as the form of the fine-structure formula and the number of quantum defect parameters required to describe the spectrum at our level of precision. The accuracy of the wave functions derived from the energy levels of the Cs atom promise to enable a host of important calculations that are not possible to do accurately now, such as those in few-body physics involving Rydberg states and their interatomic potentials. Furthermore, the accuracy of the dipole moments between different Rydberg states pushes Rydberg-atom-based sensors towards becoming a primary standard for radio frequency fields. The current state of art transition dipole moments for Cs Rydberg RF transitions 
are accurate to within 0.5 \% \cite{Simons2016}. Precise knowledge of the atomic wave functions is necessary not only to confirm this assertion but to improve the accuracy. Future work will consist of constructing a new model potential \cite{Marinescu1994} for Cs and the verification of the transition dipole moments.


\begin{acknowledgments}
This work has been supported by The National Research Council Internet of Things: Quantum Sensors Challenge program through Contract No. QSP-058-1. The calculations were supported at ITAMP by a~grant from the US National Science Foundation.
\end{acknowledgments}

\clearpage
\ifarXiv
    

\section*{Supplementary material}

# Supplemental Material:
# Ultra precise determination of Cs($nS_{1/2}$) and Cs($nD_J$) quantum defects for sensing and computing: Evaluation of core contributions

Pinrui Shen,[1] Donald Booth,[1] Chang Liu,[1] Scott Beattie,[2] Claude Marceau,[2] James P. Shaffer,[1, *]
Mariusz Pawlak,[3] and H. R. Sadeghpour[4, †]

[1] *Quantum Valley Ideas Laboratories, 485 Wes Graham Way, Waterloo, Ontario N2L 6R1, Canada*
[2] *National Research Council Canada, Ottawa, Ontario K1A 0R6, Canada*
[3] *Faculty of Chemistry, Nicolaus Copernicus University in Toruń, Gagarina 7, 87-100 Toruń, Poland*
[4] *ITAMP, Center for Astrophysics | Harvard & Smithsonian,
60 Garden St., Cambridge, Massachusetts 02138, USA*

(Dated: November 18, 2024)

## I. HISTORICAL VALUE OF THE CS IONIZATION ENERGY

Historical measurements of the Cs ionization energy are summarized in Table S1. The improvements in the measurement uncertainty are illustrated in Fig. S1. The precision of the ionization energy has been improved over the past century.

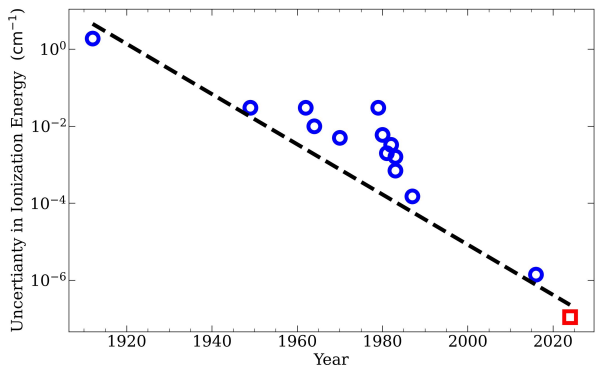

FIG. S1. Improvements in determining the ionization energy of the Cs atom over the last century. The uncertainties are reported in Refs. [1–14]. The red square represents the value reported in this work.

TABLE S1. The Cs ionization limit in cm$^{-1}$ over the years.

| Ref. | year | IE (cm$^{-1}$) |
|---|---|---|
| Bevan [1] | 1912 | 31 404.6(1.9) |
| Kratz [2] | 1949 | 31 406.54(3) |
| McNally *et al.* [3] | 1949 | 31 406.32(3) |
| Kleiman [4] | 1962 | 31 406.450(30) |
| Bockasten [5] | 1964 | 31 406.432(10) |
| Eriksson and Wenåker [6] | 1970 | 31 406.454(5) |
| Lorenzen and Niemax [7] | 1979 | 31 406.46(3) |
| Lorenzen *et al.* [8] | 1980 | 31 406.468(6) |
| Sansonetti *et al.* [9] | 1981 | 31 406.455 6(20) |
| Goy *et al.* [10] | 1982 | 31 406.467 6(33) |
| O'Sullivan and Stoicheff [11] | 1983 | 31 406.637 6(16) |
| Lorenzen and Niemax [12] | 1983 | 31 406.471 0(7) |
| Weber and Sansonetti [13] | 1987 | 31 406.467 66(15) |
| Deiglmayr *et al.* [14] | 2016 | 31 406.467 732 5(14) |
| This work | 2024 | 31 406.467 751 48(14) |

## II. ABSOLUTE FREQUENCY FITTING METHOD

Absolute frequency measurements are performed for the $nD_{5/2}, nD_{3/2}$, and $nS_{1/2}$ series. For the $nD_{3/2}$ and $nD_{5/2}$ series, the spectrum is fit to a multi-line Gaussian function,

$$f(\nu) = \sum_F (2F+1) \exp\left[-\frac{(\nu - \nu_F - \nu_0)^2}{2\sigma^2}\right], \quad (1)$$

which includes all possible hyperfine transitions weighted by the statistical factor of $(2F+1)$. $F$ represents the total atomic angular momentum. For the $6S_{1/2} \to nD_{3/2}$ transitions, the possible hyperfine quantum numbers are $F = 2$ and 3. For the $6S_{1/2} \to nD_{5/2}$ transitions, $F$ can be 1, 2, or 3. $\sigma$ is the linewidth of the spectral line, assumed to be equal for all of the hyperfine transitions, since the dominant broadening mechanism is Doppler broadening. $\nu_0$ in Eq. (1) is the frequency of the transitions from $|6S_{1/2}, F=3\rangle$ to the center of gravity of the target Rydberg state, and $\nu_F$ is the hyperfine frequency shift from the center of gravity frequency. The value of $\nu_F$ can be calculated by [15, 16]

* e-mail: jshaffer@qvil.ca
† e-mail: hrs@cfa.harvard.edu

$$\nu_F = \frac{A^*_{\text{hfs}}(n)}{2}K + B^*_{\text{hfs}}(n)\frac{\frac{3}{2}K(K+1) - 2I(I+1)J(J+1)}{4I(2I-1)J(2J-1)}$$
$$+ C^*_{\text{hfs}}(n)\frac{5K^2(K/4+1) + K[I(I+1) + J(J+1) + 3 - 3I(I+1)J(J+1)] - 5I(I+1)J(J+1)}{I(I-1)(2I-1)J(J-1)(2J-1)}, \quad (2)$$

where

$$K = F(F+1) - I(I+1) - J(J+1). \quad (3)$$

Here, $n$ is the principal quantum number, $l$ is the orbital angular momentum quantum number, $J$ is the total electron angular momentum quantum number, and $I$ is the nuclear spin quantum number. $A^*_{\text{hfs}}(n) = A_{\text{hfs}}/[n - \delta^{(l,J)}]^3$, $B^*_{\text{hfs}}(n) = B_{\text{hfs}}/[n - \delta^{(l,J)}]^3$, and $C^*_{\text{hfs}}(n) = C_{\text{hfs}}/[n - \delta^{(l,J)}]^3$ are the reduced magnetic dipole, the electric quadruple, and the magnetic octupole hyperfine coupling constants, respectively, where $\delta^{(l,J)}$ is the quantum defect. Rahaman et al. [17] performed hyperfine structure measurements for the $7D_J$ states with the quantum interference corrections and reported $A^*_{\text{hfs}}(7) = -1.7110(3)$ MHz, $B^*_{\text{hfs}}(7) = -0.027(7)$ MHz, and $C^*_{\text{hfs}}(7) = -0.02(80)$ kHz for the $7D_{5/2}$ state and $A^*_{\text{hfs}}(7) = 7.3547(8)$ MHz, $B^*_{\text{hfs}}(7) = -0.017(7)$ MHz, and $C^*_{\text{hfs}}(7) = -0.3(4)$ kHz for the $7D_{3/2}$ state. Here, we also analyze the effect of the quantum interference (QI) on the measured Rydberg spectra. In our case, the Rydberg states are detected through ionization and counted by the MCP plate. We use pulse field ionization using a fast risetime pulse applied immediately after optical excitation. The atoms are ionized before any substantial spontaneous decay of the excited Rydberg atoms occurs. To experimentally verify that the effect is minimal in our experiment, we measured the $21D_{5/2}$ spectra at different polarization angles of the second excitation step laser ($\sim 509$ nm) relative to the first excitation step laser ($\sim 852$ nm). The fit of the $21D_{5/2}$ state center frequency is plotted as a function of the polarization angle in Fig. S2. The variation in the frequency of the center of the line shape representing the energy of the Rydberg state, obtained from the fit, is $< 20$ kHz and is constant within the uncertainty of the experiment. We conclude from the experiment that we can safely ignore the QI effect in our experiment.

For lower $n$, Eq. (1) is used to extract the center of gravity frequency of the transition. An example of a measured spectrum for $6S_{1/2} \to 28D_{3/2}$ is shown as Fig. S3(a). As $n$ increases, the hyperfine frequency shift becomes smaller compared to the linewidth. A single Gaussian fitting function is used to extract the center frequency for $n > 60$, where the hyperfine splitting is calculated to be less than 10 kHz. The hyperfine frequency shift is calculated using Eq. (2) and used to correct the center frequency obtained from the Gaussian fit. The uncertainty in the hyperfine constant is also considered and taken as a source of uncertainty. For the $nS_{1/2}$ series, Eq. (1) is reduced to a single Gaussian function without

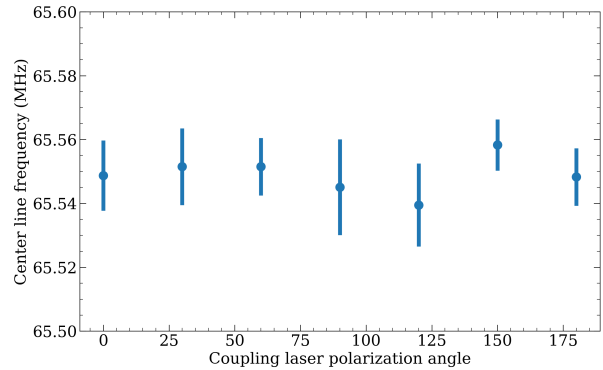

FIG. S2. Spectral line center versus the coupling laser ($\sim 509$ nm) polarization angle relative to the $\sim 852$ nm laser.

the statistical factor,

$$f(\nu) = \exp\left[-\frac{(\nu - \nu_F - \nu_0)^2}{2\sigma^2}\right], \quad (4)$$

as there is only one possible hyperfine transition, $|6S_{1/2}, F = 3\rangle \to |nS_{1/2}, F' = 3\rangle$. The magnetic dipole constant $A_{\text{hfs}}$ for the $nS_{1/2}$ series is 13.53(12) GHz [18] and the electric quadruple constant, $B_{\text{hfs}}$, and the magnetic octupole constant, $C_{\text{hfs}}$, are zero. Similar to the corrections applied for $nD_J$ series, the hyperfine splitting of the $nS_{1/2}$ series is also estimated and corrected. An example of a measured spectrum for $6S_{1/2} \to 80S_{1/2}$ is shown in Fig. S3(b). The summary of absolute frequency measurements for all three Cs series ($nS_{1/2}, nD_{3/2}$, and $nD_{5/2}$) are presented in Table S4.

## III. UNCERTAINTY ANALYSIS

In Table S4, the statistical uncertainty is given in the bracket. The dominant source contributing to the statistical uncertainty is the shot-to-shot variation in the Rydberg ion counts. The uncertainty in the number of ions is measured to be $< 10\%$ and contributes less than 10 kHz to the total statistical uncertainty.

The uncertainty caused by the ion count fluctuations increases as the spectral linewidth broadens. Therefore, efforts were made to narrow the spectral linewidth. The dominant source of line broadening is the Doppler effect. To mitigate Doppler broadening, a polarization gradient cooling (PGC) step is used and optimized to cool the atoms to $< 10$ $\mu$K. With the orthogonal ex-

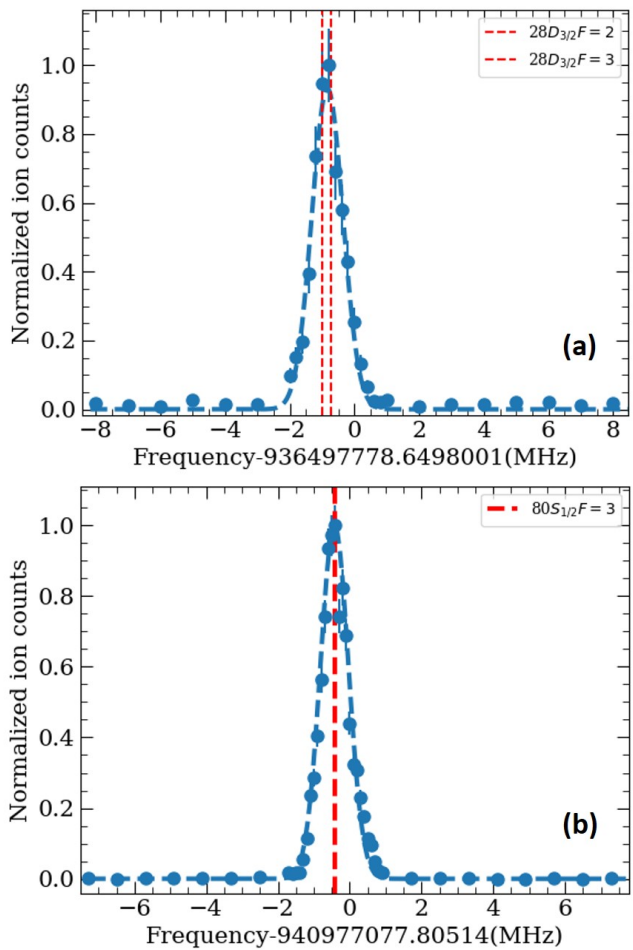

FIG. S3. Panel (a) shows an example of a measured spectrum for $6S_{1/2} \to 28D_{3/2}$. The spectrum is fit to a multi-line Gaussian function and yields a linewidth of 220 kHz. The corresponding hyperfine transitions are indicated by the red dashed lines. The lineshape is Gaussian because the dominant broadening mechanism is Doppler broadening. Panel (b) shows the spectrum for $6S_{1/2} \to 80S_{1/2}$, which is fit to a single Gaussian function (see text).

citation laser geometry, the Doppler broadening is estimated to be on the order of 200 kHz, which is close to the linewidth observed in the experiments, $\approx 250$ kHz. Other possible contributions to the observed spectral linewidth are radiative decay of the intermediate $6P_{3/2}$ state and power broadening. The radiative decay rate is reduced to $< 10$ kHz by detuning the first-step laser 90 MHz away from the $6P_{3/2}$ resonance to adiabatically eliminate the intermediate state. The power broadening is suppressed to $< 30$ kHz by setting both excitation lasers' intensities below their respective saturation intensities. Pressure broadening and transit broadening are estimated to be on the order of 100 Hz, and thus negligible.

An additional contribution to the uncertainty is the determination of the laser frequency. The two excitation lasers are both locked to a GPS steered Rb clock frequency comb which has a stability of of $1.2 \times 10^{-12}$ for a 10 s averaging time. However, the two lasers pass through AOMs before reaching the atoms. The RF signals driving the AOMs are derived from a quartz oscillator with a stability of 25 ppm. Taking into account the laser locking and the frequency shifts by the AOMs, the uncertainty in the laser frequency is $< 2$ kHz.

AC Stark shifts induced by the excitation laser fields are another source of uncertainty. Because the experiment was performed after the atoms are released from the molasses, there is no AC Stark shift from the trapping fields. The AC Stark shift induced by the $\sim 509$ nm laser is calculated to be negligible. Only the AC Stark shift on the ground state, $6S_{1/2}$, induced by the $\sim 852$ nm laser needs to be considered. To characterize this shift, we measured a series of $54S_{1/2}$ spectra at different first-step laser powers. The measured center frequencies are plotted as a function of the $\sim 852$ nm laser power, Fig. S4. The slope is found to be $(0.45 \pm 0.06)$ kHz/$\mu$W. For the experiments where we measured the Rydberg state spectra, the power of the $\sim 852$ nm laser was set to 80.0 $\mu$W, which results a $(36.0 \pm 4.8)$ kHz blue AC Stark shift. Given that the beam waist and power level of the $\sim 852$ nm laser are the same for all measurements, the AC Stark shift is applied as a shift for all Rydberg states. The shift has been applied to the values presented in Table S4 to correct the energy levels for a field-free environment. The uncertainty of the AC Stark shift, $\pm 4.8$ kHz, is taken as a contribution to the uncertainty in determining the resonance frequency, listed in Table S2.

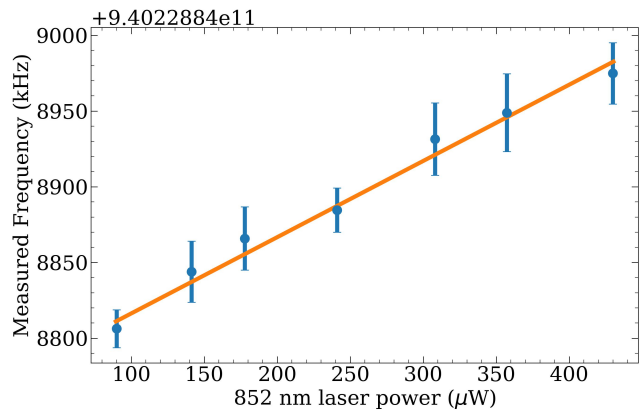

FIG. S4. AC Stark shift measurement. The absolute frequency of the $6S_{1/2} \to 54S_{1/2}$ transition is measured at different first-step laser power levels. Each data point is an average of five separate measurements and the error bar indicates the uncertainty of each data point derived from the average over the five measurements. The data is fit to a linear function (orange line), yielding a slope of $(0.45 \pm 0.06)$ kHz/$\mu$W.

Stray magnetic fields cause Zeeman shifts of the transitions. Compensation of the background magnetic field is required to obtain accurate spectroscopic measurements. The decay time of the eddy currents in the system as the MOT coils are shutdown is around 12 ms. To increase

TABLE S2. Uncertainties in the determination of the absolute frequency. The amplitude of each uncertainty is listed as the maximum possible value for any states measured in this work.

| Sources | Amplitude (kHz) |
|---|---|
| Ground state AC Stark shift | < 5.0 |
| Zeeman effect shift | < 14.3 |
| DC Stark shift | < 49.0 |
| Hyperfine splitting | < 4.0 |
| Pressure shift | < 2.8 |
| Statistical uncertianty | < 28.0 |
| Total | < 59.0 |

the decay rate of the eddy currents, we flipped the direction of the current in the MOT coils before extinguishing the magnetic field [19]. Using this approach, the decay time was shortened to less than 2.5 ms. Consequently, a wait time of 3 ms is added between the excitation stage and the extinction of the magnetic fields to reduce the magnetic fields from the eddy current. The other sources of the stray magnetic field are compensated using three pairs of Helmholtz coils placed around the MOT chamber. Microwave (MW) spectroscopy is used to determine the experimental amplitude of the total remaining magnetic field during the atom interrogation period. A similar method was used in Ref. [18] to compensate magnetic fields. The background magnetic field strength is reduced to < 2 mG during the measurement of the Rydberg state transition frequencies, which results in a maximum Zeeman splitting of around 14.3 kHz. The sensitivity to the magnetic field is similar across all Rydberg states because they have the same Landé $g$-factor. The maximum Zeeman splitting, 14.3 kHz, is treated as a common error for all Rydberg states.

DC Stark effects can cause asymmetric line broadening, as well as a frequency shift. Stray electric fields need to be compensated in order to minimize the DC Stark shift. The stray electric fields are compensated using eight electrodes around the trapped atoms. We use the $160S_{1/2}$ state as a calibration state to minimize the Stark shifts. After carefully adjusting the voltage on each electrode, the electric field is reduced to < 1 mV/cm. For the $nD_J$ series, the DC Stark effect predominately causes broadening of the spectral linewidth. Based on the polarizability of Cs $nD_J$ Rydberg states [20], the DC Stark effect introduces uncertainty that is no greater than 49 kHz for $90D_{5/2}$, while for $n = 27$, the systematic uncertainty induced by the stray electric field is no larger than 30 Hz. For the $nS_{1/2}$ series, the stray electric field primarily induces a frequency shift instead of line broadening. Due to the small polarizability of the Cs $nS_{1/2}$ states [21], the maximum DC stark effect induced frequency shift is no greater than 1.5 kHz for $90S_{1/2}$.

The pressure shift due to collisions between the Rydberg atoms and the surrounding ground state atoms is a source of uncertainty. Using the triplet $s$-wave e$^-$–Cs scattering length of $-22.7\, a_0$ [22] and the peak Cs cloud density of $2 \times 10^{10}$ cm$^{-3}$, the maximum pressure shift is estimated to be < 2.8 kHz for $n = 90$.

The main sources of the uncertainties are summarized in Table S2. For the low $n$ states, the hyperfine splitting (< 4.0 kHz) and the Zeeman splitting (< 14.3 kHz) are the key uncertainties, while for the high $n$ states, the DC Stark shift becomes dominant (< 49.0 kHz). The dominant source contributing to the statistical uncertainty is the shot-to-shot variation of the ion counts. The median statistical uncertainty is < 4.0 kHz, and the maximum statistical uncertainty is around 27.4 kHz due to one poor quality measurement.

## IV. DISCUSSION OF THE ODD POWER TERMS IN MODIFIED RITZ FORMULA

According to Ref. [23], additional odd power terms should be considered in modified Ritz equation if core polarization makes a significant contribution to the energies, because the odd power terms result from second-order dipole polarization. We fit the $nD_J$ series energy levels to modified Ritz equation with and without the odd power terms and compared the fit parameters. There are no significant changes to the values of the even quantum defects when including the odd power terms in the modified Ritz expansion. The contribution from the odd power terms is less than 0.01%. The quantum defects agree with an estimate of the core-polarization and core-penetration effects in Table II in the main text. We conclude that modified Ritz equation with only even power terms is suitable for accurately describing the energy levels of the Cs $nS_{1/2}, nP_J$, and $nD_J$ states.

## V. HIGHER ORDER ($k > 2$) QUANTUM DEFECT PARAMETERS

In this work, we report the quantum defect terms $\delta_{2k}, k=0,1,2$ up to the third order and found the higher order terms ($k > 2$) reported in previous work [12, 13] were negligible within the measurement uncertainties.

Here, we compare our reported quantum defects and prior quantum defects [13] against the observed $nS_{1/2}$ ($n = 8-90$) and $nD_{5/2}$ ($n = 8-90$) energies. The experimental data includes the measurements from Ref. [13] and the measurements in this work. We calculate the deviations characterized by the experimental uncertainty between the data and theoretical calculations using different sets of the quantum defects, as shown in Fig. S5. We find that the prior quantum defects with nonzero $k > 2$ values, obtained by fitting to observed energies for low-$n$ Rydberg levels, do not reproduce our measurements for higher-$n$ levels within uncertainty. However, the quantum defect parameters reported in this work, for high-$n$ levels with zero $k > 2$

terms *reproduce* the previous observed levels at low $n$, as described in the main text.

It is possible that the lower $n$ states are more sensitive to the higher order terms, but more precise measurements are required to extract their values. Our measurements show that the higher order $k > 2$ terms are not necessary to reproduce all the available data, including the low $n$ states, within the experimental uncertainty for all of the measurements.

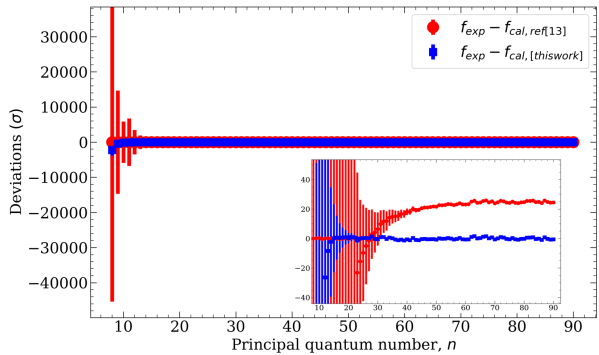

FIG. S5. Deviations between the experimental data and the energy level calculations. The experimental data consist of $nS_{1/2}$ ($n = 8-27$) measurements in Ref. [13] and $nS_{1/2}$ ($n = 28-90$) measurements in this work. Red circles represent the deviations in terms of the experimental uncertainty between the experimental data and the calculated energy levels based on the $nS_{1/2}$ quantum defects in Ref. [13]. Blue squares represent the deviations between the experimental data and the calculated energy levels based on the $nS_{1/2}$ quantum defects in this work. The error bars are the theoretical calculation uncertainties generated from the uncertainties in the quantum defects.

## VI. ANALYTICAL FORM OF THE EXPECTATION VALUE OF THE SPIN–ORBIT COUPLING FOR HYDROGENIC WAVE FUNCTIONS

We show that the dominant term of the expectation value of spin–orbit coupling for hydrogenic wave functions scales in proportion to $n^{-3}$, and the next two as $n^{-5}$ and $n^{-7}$, assuming the atomic potential to be a Coulomb potential with spherically symmetric core polarization.

The spin-orbit interaction potential is usually given by

$$V^{\mathrm{LS}}(r) = \frac{1}{2\mu^2 c^2}\left(\frac{1}{r}\frac{\mathrm{d}V}{\mathrm{d}r}\right) \mathbf{L}\cdot\mathbf{S}, \tag{5}$$

where $\mu$ is the reduced mass and $c$ is the speed of light. As an attractive potential, we take the Coulomb potential with nuclear charge $Ze$ in conjunction with the core polarization term,

$$V(r) = -\frac{1}{4\pi\varepsilon_0}\frac{Ze^2}{r} - \left(\frac{e}{4\pi\varepsilon_0}\right)^2 \frac{\alpha_c}{2r^4}, \tag{6}$$

where $\alpha_c$ is static dipole polarizability of the core. The derivative of $V(r)$ divided by $r$ is

$$\frac{1}{r}\frac{\mathrm{d}V(r)}{\mathrm{d}r} = \frac{Ze^2}{4\pi\varepsilon_0}\frac{1}{r^3} + \left(\frac{e}{4\pi\varepsilon_0}\right)^2 \frac{2\alpha_c}{r^6}. \tag{7}$$

Acting from both sides with a hydrogenic wave function, we obtain

$$\left\langle \frac{1}{r}\frac{\mathrm{d}V}{\mathrm{d}r}\right\rangle_{n,l} = \frac{Ze^2}{4\pi\varepsilon_0}\left\langle \frac{1}{r^3}\right\rangle_{n,l} + \left(\frac{e}{4\pi\varepsilon_0}\right)^2 2\alpha_c \left\langle \frac{1}{r^6}\right\rangle_{n,l}, \tag{8}$$

where

$$\left\langle \frac{1}{r^3}\right\rangle_{n,l} = \frac{Z^3}{a_\mu^3 (l+\frac{1}{2})l(l+1)}\frac{1}{n^3}, \tag{9}$$

$$\left\langle \frac{1}{r^6}\right\rangle_{n,l} = \frac{Z^6}{D}\left(A\frac{1}{n^3} + B\frac{1}{n^5} + C\frac{1}{n^7}\right), \tag{10}$$

$$A = 140, \tag{11}$$

$$B = 20[5 - 6l(l+1)], \tag{12}$$

$$C = 12l(l^2 - 1)(l+2), \tag{13}$$

$$D = a_\mu^6 l(l+1)(2l+1)[l(l+1) - 2] \\ \times [4l(l+1) - 15][4l(l+1) - 3], \tag{14}$$

and $a_\mu$ is the reduced Bohr radius, $a_\mu = \frac{m_e}{\mu}a_0$. Note that Eq. (9) holds for $l > 0$, whereas Eq. (10) for $l > 1$. The effect of the $\mathbf{L}\cdot\mathbf{S}$ term in respect of the $l, s, J, m_J$ quantum numbers is as follows

$$\langle \mathbf{L}\cdot\mathbf{S}\rangle^{(l,J)} = \frac{\hbar^2}{2}\left[J(J+1) - l(l+1) - \frac{3}{4}\right]. \tag{15}$$

The fine-structure splitting can be obtained as $\Delta_{\mathrm{fs}} \approx \langle V^{\mathrm{LS}}\rangle_{n,l,J>l} - \langle V^{\mathrm{LS}}\rangle_{n,l,J<l}$, where the dominant terms are proportional to $n^{-3}$, $n^{-5}$, and $n^{-7}$, respectively. This justifies the form of the fine-structure splitting expansion,

$$\Delta_{\mathrm{fs}} = \sum_{q=1}^{\infty} \frac{\xi_q^{(l)}}{[n - \tilde{\delta}^{(l,J=l\pm 1)}(n)]^{2q+1}}. \tag{16}$$

The weighted average quantum defect, $\tilde{\delta}^{(l,J=l\pm 1)}(n) = w^{(l,J<l)}\delta^{(l,J<l)}(n) + w^{(l,J>l)}\delta^{(l,J>l)}(n)$, comprises the quantum defects weighted by the $\mathbf{L}\cdot\mathbf{S}$ term; $w^{(l,J)} = 1 - \frac{|\langle \mathbf{L}\cdot\mathbf{S}\rangle^{(l,J)}|}{\hbar^2(l+1/2)}$.

## VII. NUMERICAL CALCULATIONS OF WAVE FUNCTIONS AND REDUCED ELECTRIC-DIPOLE MATRIX ELEMENTS

The Hamiltonian describing the valence electron in the field of an alkali-metal positive-ion core reads (in a.u.)

$$\hat{H} = -\frac{1}{2\mu}\nabla^2 + V_l(r) + V_l^{\mathrm{LS}}(r), \tag{17}$$

where $\mu$ is the reduced mass, $V_l(r)$ is the model interaction potential [24], and $V_l^{\text{LS}}(r)$ is the spin–orbit coupling [25]. The $l$-dependent effective potential, $V_l(r)$, including a polarization term is (in a.u.)

$$V_l(r) = -\frac{1+(Z-1)\exp\left[-a_1^{(l)}r\right]}{r} \\ + \left(a_3^{(l)} + a_4^{(l)}r\right)\exp\left[-a_2^{(l)}r\right] \\ - \frac{\alpha_c}{2r^4}\left(1 - \exp\left[-\left(r/r_c^{(l)}\right)^6\right]\right), \quad (18)$$

where $Z$ is the nuclear charge number and $\alpha_c$ is the static dipole polarizability. Five parameters ($r_c^{(l)}$, $a_1^{(l)}$, $a_2^{(l)}$, $a_3^{(l)}$, and $a_4^{(l)}$) are given for different alkali-metal atoms in Ref. [24]. The next term in the Hamiltonian, describing the spin–orbit interaction, could be expressed in the non-trivial form (in a.u.) [25]

$$V_l^{\text{LS}}(r) = \frac{1}{2\mu^2c^2}\left(\frac{1}{r}\frac{dV_l(r)}{dr}\right)\mathbf{L}\cdot\mathbf{S}\left(1 - \frac{1}{2\mu^2c^2}V_l(r)\right)^{-2}. \quad (19)$$

The last part of $V_l^{\text{LS}}$ guarantees proper behavior of wave functions at short range.

We calculated the wave functions of the time-independent Schrödinger equation by numerical integration utilizing Numerov's method with energies determined from our quantum defects. We set the integration interval as $[0.01, 2n(n+15)]$ (in a.u.), divided into $4\times 10^6$ equal segments. All obtained functions have attained the appropriate behavior close to the origin as well as asymptotic behavior. We calculated (within composite Simpson's rule) the expectation values of $r$, $V_l^{\text{LS}}$, and each term of $V_l$. Subsequently, we estimated the numerical $\Delta_{\text{fs}}$ for the $nD_J$ series by $\langle V_{l=2}^{\text{LS}}\rangle_{n,l=2,J=5/2} - \langle V_{l=2}^{\text{LS}}\rangle_{n,l=2,J=3/2}$.

The model potential is viewed as a combination interaction of core penetration and core polarization. The first two terms in Eq. (18) are usually considered as the core penetration dominant term, $V_{\text{pen}}(r)$, while the third term is the result of core polarization, $V_{\text{pol}}(r)$. Then, the energy contributions from core polarization and core penetration are $\langle V_{\text{pol}}\rangle$ and $\langle V_{\text{pen}}\rangle$, respectively.

Moreover, based on our results and quantum defects for the $nP_J$ states from Ref. [14], we determined the reduced electric-dipole matrix elements for $n'P_{J'} \leftrightarrow nD_J$ transitions according to

$$\langle J'||r||J\rangle = (-1)^{l+J+s'+1}\sqrt{(2J'+1)(2J+1)}\delta_{s',s} \\ \times \begin{Bmatrix} l' & J' & s' \\ J & l & 1 \end{Bmatrix}\langle l'||r||l\rangle, \quad (20)$$

where

$$\langle l'||r||l\rangle = (-1)^{l'}\sqrt{(2l'+1)(2l+1)} \\ \times \begin{pmatrix} l' & 1 & l \\ 0 & 0 & 0 \end{pmatrix}\langle n'l'J'|r|nlJ\rangle. \quad (21)$$

The results for low-lying states are compared with the computed values of Safronova *et al.* [26] and for highly excited states with the results from ARC [27], see Table S3.

TABLE S3. The reduced electric-dipole matrix elements (in a.u.).

| Transition | Other works | This work |
|---|---|---|
| $7D_{3/2} \leftrightarrow 7P_{1/2}$ | 6.6(2) [26] | 6.5814 |
| $7D_{3/2} \leftrightarrow 8P_{1/2}$ | 32.0(1) [26] | 31.9943 |
| $7D_{3/2} \leftrightarrow 9P_{1/2}$ | 9.0(2) [26] | 8.8867 |
| $7D_{3/2} \leftrightarrow 10P_{1/2}$ | 2.86(8) [26] | 2.8524 |
| $7D_{3/2} \leftrightarrow 11P_{1/2}$ | 1.55(4) [26] | 1.5506 |
| $7D_{3/2} \leftrightarrow 12P_{1/2}$ | 1.03(3) [26] | 1.0268 |
| $7D_{3/2} \leftrightarrow 7P_{3/2}$ | 3.3(1) [26] | 3.3328 |
| $7D_{3/2} \leftrightarrow 8P_{3/2}$ | 14.35(5) [26] | 14.3587 |
| $7D_{3/2} \leftrightarrow 9P_{3/2}$ | 3.56(9) [26] | 3.5040 |
| $7D_{3/2} \leftrightarrow 10P_{3/2}$ | 1.16(4) [26] | 1.1603 |
| $7D_{3/2} \leftrightarrow 11P_{3/2}$ | 0.64(2) [26] | 0.6354 |
| $7D_{3/2} \leftrightarrow 12P_{3/2}$ | 0.42(1) [26] | 0.4219 |
| $7D_{5/2} \leftrightarrow 7P_{3/2}$ | 9.6(3) [26] | 9.6960 |
| $7D_{5/2} \leftrightarrow 8P_{3/2}$ | 43.2(1) [26] | 43.2245 |
| $7D_{5/2} \leftrightarrow 9P_{3/2}$ | 11.1(2) [26] | 10.9461 |
| $7D_{5/2} \leftrightarrow 10P_{3/2}$ | 3.6(1) [26] | 3.5939 |
| $7D_{5/2} \leftrightarrow 11P_{3/2}$ | 1.97(6) [26] | 1.9640 |
| $7D_{5/2} \leftrightarrow 12P_{3/2}$ | 1.31(4) [26] | 1.3030 |
| $8D_{3/2} \leftrightarrow 9P_{1/2}$ | 49.3(1) [26] | 49.2834 |
| $8D_{3/2} \leftrightarrow 9P_{3/2}$ | 22.14(7) [26] | 22.1317 |
| $8D_{5/2} \leftrightarrow 9P_{3/2}$ | 66.6(2) [26] | 66.5588 |
| $9D_{3/2} \leftrightarrow 10P_{1/2}$ | 70.0(1) [26] | 69.9676 |
| $9D_{3/2} \leftrightarrow 10P_{3/2}$ | 31.45(8) [26] | 31.4347 |
| $9D_{5/2} \leftrightarrow 10P_{3/2}$ | 94.5(2) [26] | 94.4738 |
| $10D_{3/2} \leftrightarrow 11P_{1/2}$ | 94.1(2) [26] | 94.0566 |
| $10D_{3/2} \leftrightarrow 11P_{3/2}$ | 42.29(9) [26] | 42.2721 |
| $10D_{5/2} \leftrightarrow 11P_{3/2}$ | 127.0(2) [26] | 126.9827 |
| $11D_{3/2} \leftrightarrow 12P_{1/2}$ | 121.6(2) [26] | 121.5538 |
| $11D_{3/2} \leftrightarrow 12P_{3/2}$ | 54.66(9) [26] | 54.6452 |
| $11D_{5/2} \leftrightarrow 12P_{3/2}$ | 164.1(2) [26] | 164.0899 |
| $30D_{3/2} \leftrightarrow 30P_{1/2}$ | 164.5663 [27] | 164.5625 |
| $30D_{3/2} \leftrightarrow 30P_{3/2}$ | 89.8979 [27] | 89.8519 |
| $30D_{5/2} \leftrightarrow 30P_{3/2}$ | 255.9199 [27] | 255.7829 |
| $40D_{3/2} \leftrightarrow 40P_{1/2}$ | 297.6955 [27] | 297.6939 |
| $40D_{3/2} \leftrightarrow 40P_{3/2}$ | 163.4144 [27] | 163.3215 |
| $40D_{5/2} \leftrightarrow 40P_{3/2}$ | 464.5782 [27] | 464.2992 |
| $50D_{3/2} \leftrightarrow 50P_{1/2}$ | 469.9097 [27] | 469.9095 |
| $50D_{3/2} \leftrightarrow 50P_{3/2}$ | 258.6886 [27] | 258.5334 |
| $50D_{5/2} \leftrightarrow 50P_{3/2}$ | 734.8516 [27] | 734.3844 |
| $60D_{3/2} \leftrightarrow 60P_{1/2}$ | 681.2107 [27] | 681.2114 |
| $60D_{3/2} \leftrightarrow 60P_{3/2}$ | 375.7211 [27] | 375.4885 |
| $60D_{5/2} \leftrightarrow 60P_{3/2}$ | 1066.7426 [27] | 1066.0418 |
| $70D_{3/2} \leftrightarrow 70P_{1/2}$ | 931.5993 [27] | 931.6004 |
| $70D_{3/2} \leftrightarrow 70P_{3/2}$ | 514.5122 [27] | 514.1874 |
| $70D_{5/2} \leftrightarrow 70P_{3/2}$ | 1460.2520 [27] | 1459.2724 |
| $80D_{3/2} \leftrightarrow 80P_{1/2}$ | 1221.0759 [27] | 1221.0769 |
| $80D_{3/2} \leftrightarrow 80P_{3/2}$ | 675.0622 [27] | 674.6302 |
| $80D_{5/2} \leftrightarrow 80P_{3/2}$ | 1915.3803 [27] | 1914.0768 |

TABLE S4. Absolute frequency measurements and the statistical uncertainties of transitions from the $|6S_{1/2}, F = 3\rangle$ hyperfine ground state to the center of gravity of the $nS_{1/2}$, $nD_{5/2}$, $nD_{3/2}$ states. The statistical uncertainties are presented in the brackets.

| $n$ | $\nu_{S_{1/2}}$(kHz) | $\nu_{D_{5/2}}$(kHz) | $\nu_{D_{3/2}}$(kHz) |
|---|---|---|---|
| 21 | — | 931969904159.4(6.2) | 931960462840.3(16.9) |
| 22 | — | 932925420315.1(9.9) | 932917355120.8(27.0) |
| 23 | 932386112252.8(10.9) | 933744762535.1(6.1) | 933737818429.9(5.3) |
| 24 | 933281572697.2(9.1) | 934452628487.4(8.2) | 934446607025.4(5.7) |
| 25 | 934051877594.8(6.0) | 935068360508.0(5.3) | 935063105302.8(7.6) |
| 26 | 934719309793.5(6.2) | 935607280846.3(4.3) | 935602667250.9(11.4) |
| 27 | 935301404781.5(5.8) | 936081653844.5(6.7) | 936077581723.7(25.5) |
| 28 | 935812114821.2(6.6) | 936501390829.7(11.6) | 936497778401.0(9.4) |
| 29 | 936262652614.1(6.6) | 936874570954.1(11.0) | 936871351515.9(14.4) |
| 30 | 936662113134.8(23.2) | 937207833091.9(3.5) | 937204951761.9(9.0) |
| 31 | 937017935503.2(7.8) | 937506674118.6(7.4) | 937504085293.8(17.5) |
| 32 | 937336251966.1(6.8) | 937775676305.4(12.9) | 937773341520.9(27.4) |
| 33 | 937622154361.1(6.6) | 938018683599.5(12.0) | 938016570610.9(21.0) |
| 34 | 937879899105.2(5.8) | 938238940313.9(6.9) | 938237022105.7(20.4) |
| 35 | 938113065775.6(5.9) | 938439200875.3(7.1) | 938437453971.2(10.3) |
| 36 | 938324682925.2(7.5) | 938621813939.2(11.5) | 938620218885.2(18.0) |
| 37 | 938517326596.3(6.1) | 938788794105.1(6.1) | 938787333466.2(11.6) |
| 38 | 938693199099.5(13.8) | 938941875772.7(8.4) | 938940534869.6(13.0) |
| 39 | 938854192054.4(6.7) | 939082559923.6(6.3) | 939081326162.6(11.3) |
| 40 | 939001938005.6(4.9) | 939212149979.8(5.3) | 939211012223.6(11.8) |
| 41 | 939137851248.8(5.5) | 939331782115.4(3.3) | 939330730785.5(6.5) |
| 42 | 939263162645.1(4.3) | 939442451581.3(3.2) | 939441477865.5(6.1) |
| 43 | 939378946543.6(4.4) | 939545031475.7(2.8) | 939544128149.5(4.5) |
| 44 | 939486145108.6(5.6) | 939640291542.0(3.8) | 939639452038.1(4.0) |
| 45 | 939585586643.1(5.6) | 939728912110.0(2.2) | 939728130311.4(4.1) |
| 46 | 939678001804.1(3.9) | 939811496008.9(2.3) | 939810766742.1(3.9) |
| 47 | 939764037301.9(4.5) | 939888579281.8(4.5) | 939887898202.6(4.5) |
| 48 | 939844267171.4(3.3) | 939960639986.8(3.9) | 939960002633.6(6.9) |
| 49 | 939919202219.9(4.1) | 940028105046.6(2.7) | 940027507933.1(5.2) |
| 50 | 939989298297.8(3.1) | 940091356980.5(2.3) | 940090796847.7(6.3) |
| 51 | 940054963356.3(3.7) | 940150739757.1(3.0) | 940150213503.5(4.0) |
| 52 | 940116562993.4(3.2) | 940206562229.9(3.3) | 940206067286.4(5.1) |
| 53 | 940174426104.1(3.2) | 940259103796.2(3.9) | 940258637730.8(4.0) |
| 54 | 940228848894.2(3.3) | 940308616656.9(2.8) | 940308176949.6(4.0) |
| 55 | 940280098714.1(2.8) | 940355328595.1(2.5) | 940354913714.3(3.4) |
| 56 | 940328417535.4(2.6) | 940399447726.2(2.7) | 940399055410.9(4.0) |
| 57 | 940374024757.6(2.6) | 940441161862.7(2.7) | 940440790798.0(3.7) |
| 58 | 940417119508.6(2.8) | 940480642898.3(3.0) | 940480291683.7(3.7) |
| 59 | 940457882873.4(2.7) | 940518047613.3(2.9) | 940517714586.4(4.0) |
| 60 | 940496480189.1(2.9) | 940553518755.5(2.8) | 940553202814.9(3.7) |
| 61 | 940533062211.7(3.9) | 940587187563.1(2.9) | 940586887573.7(4.0) |
| 62 | 940567766937.4(2.6) | 940619174003.0(2.9) | 940618888798.3(4.0) |
| 63 | 940600720478.7(2.8) | 940649588356.1(3.1) | 940649317128.1(4.6) |
| 64 | 940632038624.7(2.5) | 940678532080.7(3.3) | 940678273721.2(6.8) |
| 65 | 940661828075.4(2.9) | 940706098240.3(3.0) | 940705852154.6(3.6) |
| 66 | 940690186672.2(2.4) | 940732373107.7(2.8) | 940732138396.3(3.5) |
| 67 | 940717204520.9(2.7) | 940757435910.2(2.9) | 940757211965.9(3.4) |
| 68 | 940742964737.3(2.8) | 940781360273.0(3.3) | 940781146543.5(4.9) |
| 69 | 940767544436.6(2.9) | 940804213858.0(3.3) | 940804009611.6(4.3) |
| 70 | 940791014390.2(2.6) | 940826059993.4(2.9) | 940825864543.3(4.0) |
| 71 | 940813440688.8(2.9) | 940846956605.6(3.2) | 940846769709.6(5.1) |
| 72 | 940834884192.3(3.2) | 940866958354.9(2.9) | 940866779386.3(4.3) |
| 73 | 940855401394.3(3.0) | 940886115318.8(3.1) | 940885943816.9(4.5) |
| 74 | 940875044929.6(3.0) | 940904474457.7(3.0) | 940904310172.5(3.9) |
| 75 | 940893863785.4(2.6) | 940922079455.2(3.8) | 940921921920.8(4.3) |
| 76 | 940911903524.6(2.5) | 940938971360.0(3.5) | 940938819935.9(4.9) |
| 77 | 940929206377.2(3.4) | 940955187877.3(3.0) | 940955042412.3(4.1) |
| 78 | 940945812091.5(4.2) | 940970764457.6(2.8) | 940970624780.2(5.0) |
| 79 | 940961757637.3(2.8) | 940985734322.3(3.1) | 940985600177.7(4.7) |
| 80 | 940977077468.5(3.0) | 941000129023.2(2.8) | 940999999914.6(5.0) |
| 81 | 940991803979.2(3.1) | 941013977102.4(4.3) | 941013852909.6(4.1) |
| 82 | 941005967315.5(3.3) | 941027306402.3(2.8) | 941027186804.4(4.4) |
| 83 | 941019595894.5(3.0) | 941040141871.0(3.0) | 941040026757.3(4.0) |
| 84 | 941032716195.4(3.3) | 941052508233.8(3.6) | 941052397262.3(6.7) |
| 85 | 941045353362.2(2.8) | 941064427739.9(3.8) | 941064320777.0(4.3) |
| 86 | 941057530672.1(4.0) | 941075921921.6(2.9) | 941075818628.5(3.8) |
| 87 | 941069270411.2(3.2) | 941087010435.1(3.6) | 941086910744.2(3.9) |
| 88 | 941080592943.8(3.0) | 941097712195.3(3.0) | 941097616114.7(5.5) |
| 89 | 941091518016.1(3.5) | 941108045350.2(4.5) | 941107952468.8(6.4) |
| 90 | 941102064029.3(3.1) | 941118026228.4(4.7) | 941117936446.0(7.2) |


\fi


\begin{thebibliography}{39}%
\makeatletter
\providecommand \@ifxundefined [1]{%
 \@ifx{#1\undefined}
}%
\providecommand \@ifnum [1]{%
 \ifnum #1\expandafter \@firstoftwo
 \else \expandafter \@secondoftwo
 \fi
}%
\providecommand \@ifx [1]{%
 \ifx #1\expandafter \@firstoftwo
 \else \expandafter \@secondoftwo
 \fi
}%
\providecommand \natexlab [1]{#1}%
\providecommand \enquote  [1]{``#1''}%
\providecommand \bibnamefont  [1]{#1}%
\providecommand \bibfnamefont [1]{#1}%
\providecommand \citenamefont [1]{#1}%
\providecommand \href@noop [0]{\@secondoftwo}%
\providecommand \href [0]{\begingroup \@sanitize@url \@href}%
\providecommand \@href[1]{\@@startlink{#1}\@@href}%
\providecommand \@@href[1]{\endgroup#1\@@endlink}%
\providecommand \@sanitize@url [0]{\catcode `\\12\catcode `\$12\catcode
  `\&12\catcode `\#12\catcode `\^12\catcode `\_12\catcode `\%12\relax}%
\providecommand \@@startlink[1]{}%
\providecommand \@@endlink[0]{}%
\providecommand \url  [0]{\begingroup\@sanitize@url \@url }%
\providecommand \@url [1]{\endgroup\@href {#1}{\urlprefix }}%
\providecommand \urlprefix  [0]{URL }%
\providecommand \Eprint [0]{\href }%
\providecommand \doibase [0]{https://doi.org/}%
\providecommand \selectlanguage [0]{\@gobble}%
\providecommand \bibinfo  [0]{\@secondoftwo}%
\providecommand \bibfield  [0]{\@secondoftwo}%
\providecommand \translation [1]{[#1]}%
\providecommand \BibitemOpen [0]{}%
\providecommand \bibitemStop [0]{}%
\providecommand \bibitemNoStop [0]{.\EOS\space}%
\providecommand \EOS [0]{\spacefactor3000\relax}%
\providecommand \BibitemShut  [1]{\csname bibitem#1\endcsname}%
\let\auto@bib@innerbib\@empty
\bibitem [{\citenamefont {Safronova}\ \emph {et~al.}(2018)\citenamefont
  {Safronova}, \citenamefont {Budker}, \citenamefont {DeMille}, \citenamefont
  {Kimball}, \citenamefont {Derevianko},\ and\ \citenamefont
  {Clark}}]{Safronova2018}%
  \BibitemOpen
  \bibfield  {author} {\bibinfo {author} {\bibfnamefont {M.~S.}\ \bibnamefont
  {Safronova}}, \bibinfo {author} {\bibfnamefont {D.}~\bibnamefont {Budker}},
  \bibinfo {author} {\bibfnamefont {D.}~\bibnamefont {DeMille}}, \bibinfo
  {author} {\bibfnamefont {D.~F.~J.}\ \bibnamefont {Kimball}}, \bibinfo
  {author} {\bibfnamefont {A.}~\bibnamefont {Derevianko}},\ and\ \bibinfo
  {author} {\bibfnamefont {C.~W.}\ \bibnamefont {Clark}},\ }\bibfield  {title}
  {\bibinfo {title} {Search for new physics with atoms and molecules},\ }\href
  {https://doi.org/10.1103/RevModPhys.90.025008} {\bibfield  {journal}
  {\bibinfo  {journal} {Rev. Mod. Phys.}\ }\textbf {\bibinfo {volume} {90}},\
  \bibinfo {pages} {025008} (\bibinfo {year} {2018})}\BibitemShut {NoStop}%
\bibitem [{\citenamefont {Myers}(2019)}]{Myers2019}%
  \BibitemOpen
  \bibfield  {author} {\bibinfo {author} {\bibfnamefont {E.~G.}\ \bibnamefont
  {Myers}},\ }\bibfield  {title} {\bibinfo {title} {High-precision atomic mass
  measurements for fundamental constants},\ }\href
  {https://doi.org/10.3390/atoms7010037} {\bibfield  {journal} {\bibinfo
  {journal} {Atoms}\ }\textbf {\bibinfo {volume} {7}},\ \bibinfo {pages} {37}
  (\bibinfo {year} {2019})}\BibitemShut {NoStop}%
\bibitem [{\citenamefont {Cronin}\ \emph {et~al.}(2009)\citenamefont {Cronin},
  \citenamefont {Schmiedmayer},\ and\ \citenamefont
  {Pritchard}}]{Pritchard2009}%
  \BibitemOpen
  \bibfield  {author} {\bibinfo {author} {\bibfnamefont {A.~D.}\ \bibnamefont
  {Cronin}}, \bibinfo {author} {\bibfnamefont {J.}~\bibnamefont
  {Schmiedmayer}},\ and\ \bibinfo {author} {\bibfnamefont {D.~E.}\ \bibnamefont
  {Pritchard}},\ }\bibfield  {title} {\bibinfo {title} {Optics and
  interferometry with atoms and molecules},\ }\href
  {https://doi.org/10.1103/RevModPhys.81.1051} {\bibfield  {journal} {\bibinfo
  {journal} {Rev. Mod. Phys.}\ }\textbf {\bibinfo {volume} {81}},\ \bibinfo
  {pages} {1051} (\bibinfo {year} {2009})}\BibitemShut {NoStop}%
\bibitem [{\citenamefont {Weisskopf}\ \emph {et~al.}(1968)\citenamefont
  {Weisskopf}, \citenamefont {Carrico}, \citenamefont {Gould}, \citenamefont
  {Lipworth},\ and\ \citenamefont {Stein}}]{Weisskopf1968}%
  \BibitemOpen
  \bibfield  {author} {\bibinfo {author} {\bibfnamefont {M.~C.}\ \bibnamefont
  {Weisskopf}}, \bibinfo {author} {\bibfnamefont {J.~P.}\ \bibnamefont
  {Carrico}}, \bibinfo {author} {\bibfnamefont {H.}~\bibnamefont {Gould}},
  \bibinfo {author} {\bibfnamefont {E.}~\bibnamefont {Lipworth}},\ and\
  \bibinfo {author} {\bibfnamefont {T.~S.}\ \bibnamefont {Stein}},\ }\bibfield
  {title} {\bibinfo {title} {Electric dipole moment of the cesium atom. {A} new
  upper limit to the electric dipole moment of the electron},\ }\href
  {https://doi.org/10.1103/PhysRevLett.21.1645} {\bibfield  {journal} {\bibinfo
   {journal} {Phys. Rev. Lett.}\ }\textbf {\bibinfo {volume} {21}},\ \bibinfo
  {pages} {1645} (\bibinfo {year} {1968})}\BibitemShut {NoStop}%
\bibitem [{\citenamefont {Murthy}\ \emph {et~al.}(1989)\citenamefont {Murthy},
  \citenamefont {{Krause, Jr.}}, \citenamefont {Li},\ and\ \citenamefont
  {Hunter}}]{Murthy1989}%
  \BibitemOpen
  \bibfield  {author} {\bibinfo {author} {\bibfnamefont {S.~A.}\ \bibnamefont
  {Murthy}}, \bibinfo {author} {\bibfnamefont {D.}~\bibnamefont {{Krause,
  Jr.}}}, \bibinfo {author} {\bibfnamefont {Z.~L.}\ \bibnamefont {Li}},\ and\
  \bibinfo {author} {\bibfnamefont {L.~R.}\ \bibnamefont {Hunter}},\ }\bibfield
   {title} {\bibinfo {title} {New limits on the electron electric dipole moment
  from cesium},\ }\href {https://doi.org/10.1103/PhysRevLett.63.965} {\bibfield
   {journal} {\bibinfo  {journal} {Phys. Rev. Lett.}\ }\textbf {\bibinfo
  {volume} {63}},\ \bibinfo {pages} {965} (\bibinfo {year} {1989})}\BibitemShut
  {NoStop}%
\bibitem [{\citenamefont {Bernreuther}\ and\ \citenamefont
  {Suzuki}(1991)}]{Bernreuther1991}%
  \BibitemOpen
  \bibfield  {author} {\bibinfo {author} {\bibfnamefont {W.}~\bibnamefont
  {Bernreuther}}\ and\ \bibinfo {author} {\bibfnamefont {M.}~\bibnamefont
  {Suzuki}},\ }\bibfield  {title} {\bibinfo {title} {The electric dipole moment
  of the electron},\ }\href {https://doi.org/10.1103/RevModPhys.63.313}
  {\bibfield  {journal} {\bibinfo  {journal} {Rev. Mod. Phys.}\ }\textbf
  {\bibinfo {volume} {63}},\ \bibinfo {pages} {313} (\bibinfo {year}
  {1991})}\BibitemShut {NoStop}%
\bibitem [{\citenamefont {Chin}\ \emph {et~al.}(2001)\citenamefont {Chin},
  \citenamefont {Leiber}, \citenamefont {Vuleti\ifmmode~\acute{c}\else
  \'{c}\fi{}}, \citenamefont {Kerman},\ and\ \citenamefont {Chu}}]{Chin2001}%
  \BibitemOpen
  \bibfield  {author} {\bibinfo {author} {\bibfnamefont {C.}~\bibnamefont
  {Chin}}, \bibinfo {author} {\bibfnamefont {V.}~\bibnamefont {Leiber}},
  \bibinfo {author} {\bibfnamefont {V.}~\bibnamefont
  {Vuleti\ifmmode~\acute{c}\else \'{c}\fi{}}}, \bibinfo {author} {\bibfnamefont
  {A.~J.}\ \bibnamefont {Kerman}},\ and\ \bibinfo {author} {\bibfnamefont
  {S.}~\bibnamefont {Chu}},\ }\bibfield  {title} {\bibinfo {title} {Measurement
  of an electron's electric dipole moment using {C}s atoms trapped in optical
  lattices},\ }\href {https://doi.org/10.1103/PhysRevA.63.033401} {\bibfield
  {journal} {\bibinfo  {journal} {Phys. Rev. A}\ }\textbf {\bibinfo {volume}
  {63}},\ \bibinfo {pages} {033401} (\bibinfo {year} {2001})}\BibitemShut
  {NoStop}%
\bibitem [{\citenamefont {Wilpers}\ \emph {et~al.}(2002)\citenamefont
  {Wilpers}, \citenamefont {Binnewies}, \citenamefont {Degenhardt},
  \citenamefont {Sterr}, \citenamefont {Helmcke},\ and\ \citenamefont
  {Riehle}}]{Wilpers2002}%
  \BibitemOpen
  \bibfield  {author} {\bibinfo {author} {\bibfnamefont {G.}~\bibnamefont
  {Wilpers}}, \bibinfo {author} {\bibfnamefont {T.}~\bibnamefont {Binnewies}},
  \bibinfo {author} {\bibfnamefont {C.}~\bibnamefont {Degenhardt}}, \bibinfo
  {author} {\bibfnamefont {U.}~\bibnamefont {Sterr}}, \bibinfo {author}
  {\bibfnamefont {J.}~\bibnamefont {Helmcke}},\ and\ \bibinfo {author}
  {\bibfnamefont {F.}~\bibnamefont {Riehle}},\ }\bibfield  {title} {\bibinfo
  {title} {Optical clock with ultracold neutral atoms},\ }\href
  {https://doi.org/10.1103/PhysRevLett.89.230801} {\bibfield  {journal}
  {\bibinfo  {journal} {Phys. Rev. Lett.}\ }\textbf {\bibinfo {volume} {89}},\
  \bibinfo {pages} {230801} (\bibinfo {year} {2002})}\BibitemShut {NoStop}%
\bibitem [{\citenamefont {Pan}\ \emph {et~al.}(2020)\citenamefont {Pan},
  \citenamefont {Arora}, \citenamefont {Yu}, \citenamefont {Sahoo},\ and\
  \citenamefont {Chen}}]{Pan2020}%
  \BibitemOpen
  \bibfield  {author} {\bibinfo {author} {\bibfnamefont {D.}~\bibnamefont
  {Pan}}, \bibinfo {author} {\bibfnamefont {B.}~\bibnamefont {Arora}}, \bibinfo
  {author} {\bibfnamefont {Y.-m.}\ \bibnamefont {Yu}}, \bibinfo {author}
  {\bibfnamefont {B.~K.}\ \bibnamefont {Sahoo}},\ and\ \bibinfo {author}
  {\bibfnamefont {J.}~\bibnamefont {Chen}},\ }\bibfield  {title} {\bibinfo
  {title} {Optical-lattice-based {C}s active clock with a continual
  superradiant lasing signal},\ }\href
  {https://doi.org/10.1103/PhysRevA.102.041101} {\bibfield  {journal} {\bibinfo
   {journal} {Phys. Rev. A}\ }\textbf {\bibinfo {volume} {102}},\ \bibinfo
  {pages} {041101} (\bibinfo {year} {2020})}\BibitemShut {NoStop}%
\bibitem [{\citenamefont {Ludlow}\ \emph {et~al.}(2015)\citenamefont {Ludlow},
  \citenamefont {Boyd}, \citenamefont {Ye}, \citenamefont {Peik},\ and\
  \citenamefont {Schmidt}}]{Ludlow2015}%
  \BibitemOpen
  \bibfield  {author} {\bibinfo {author} {\bibfnamefont {A.~D.}\ \bibnamefont
  {Ludlow}}, \bibinfo {author} {\bibfnamefont {M.~M.}\ \bibnamefont {Boyd}},
  \bibinfo {author} {\bibfnamefont {J.}~\bibnamefont {Ye}}, \bibinfo {author}
  {\bibfnamefont {E.}~\bibnamefont {Peik}},\ and\ \bibinfo {author}
  {\bibfnamefont {P.~O.}\ \bibnamefont {Schmidt}},\ }\bibfield  {title}
  {\bibinfo {title} {Optical atomic clocks},\ }\href
  {https://doi.org/10.1103/RevModPhys.87.637} {\bibfield  {journal} {\bibinfo
  {journal} {Rev. Mod. Phys.}\ }\textbf {\bibinfo {volume} {87}},\ \bibinfo
  {pages} {637} (\bibinfo {year} {2015})}\BibitemShut {NoStop}%
\bibitem [{\citenamefont {Sharma}\ \emph {et~al.}(2022)\citenamefont {Sharma},
  \citenamefont {Kolkowitz},\ and\ \citenamefont {Saffman}}]{Sharma2022}%
  \BibitemOpen
  \bibfield  {author} {\bibinfo {author} {\bibfnamefont {A.}~\bibnamefont
  {Sharma}}, \bibinfo {author} {\bibfnamefont {S.}~\bibnamefont {Kolkowitz}},\
  and\ \bibinfo {author} {\bibfnamefont {M.}~\bibnamefont {Saffman}},\
  }\bibfield  {title} {\bibinfo {title} {Analysis of a cesium lattice optical
  clock},\ }\href@noop {} {\  (\bibinfo {year} {2022})},\ \Eprint
  {https://arxiv.org/abs/2203.08708} {arXiv:2203.08708 [quant-ph]} \BibitemShut
  {NoStop}%
\bibitem [{\citenamefont {Bothwell}\ \emph {et~al.}(2022)\citenamefont
  {Bothwell}, \citenamefont {Kennedy}, \citenamefont {Aeppli}, \citenamefont
  {Kedar}, \citenamefont {Robinson}, \citenamefont {Oelker}, \citenamefont
  {Staron},\ and\ \citenamefont {Ye}}]{Bothwell2022}%
  \BibitemOpen
  \bibfield  {author} {\bibinfo {author} {\bibfnamefont {T.}~\bibnamefont
  {Bothwell}}, \bibinfo {author} {\bibfnamefont {C.~J.}\ \bibnamefont
  {Kennedy}}, \bibinfo {author} {\bibfnamefont {A.}~\bibnamefont {Aeppli}},
  \bibinfo {author} {\bibfnamefont {D.}~\bibnamefont {Kedar}}, \bibinfo
  {author} {\bibfnamefont {J.~M.}\ \bibnamefont {Robinson}}, \bibinfo {author}
  {\bibfnamefont {E.}~\bibnamefont {Oelker}}, \bibinfo {author} {\bibfnamefont
  {A.}~\bibnamefont {Staron}},\ and\ \bibinfo {author} {\bibfnamefont
  {J.}~\bibnamefont {Ye}},\ }\bibfield  {title} {\bibinfo {title} {Resolving
  the gravitational redshift across a millimetre-scale atomic sample},\ }\href
  {https://doi.org/10.1038/s41586-021-04349-7} {\bibfield  {journal} {\bibinfo
  {journal} {Nature}\ }\textbf {\bibinfo {volume} {602}},\ \bibinfo {pages}
  {420} (\bibinfo {year} {2022})}\BibitemShut {NoStop}%
\bibitem [{\citenamefont {Graham}\ \emph {et~al.}(2019)\citenamefont {Graham},
  \citenamefont {Kwon}, \citenamefont {Grinkemeyer}, \citenamefont {Marra},
  \citenamefont {Jiang}, \citenamefont {Lichtman}, \citenamefont {Sun},
  \citenamefont {Ebert},\ and\ \citenamefont {Saffman}}]{Graham2019}%
  \BibitemOpen
  \bibfield  {author} {\bibinfo {author} {\bibfnamefont {T.~M.}\ \bibnamefont
  {Graham}}, \bibinfo {author} {\bibfnamefont {M.}~\bibnamefont {Kwon}},
  \bibinfo {author} {\bibfnamefont {B.}~\bibnamefont {Grinkemeyer}}, \bibinfo
  {author} {\bibfnamefont {Z.}~\bibnamefont {Marra}}, \bibinfo {author}
  {\bibfnamefont {X.}~\bibnamefont {Jiang}}, \bibinfo {author} {\bibfnamefont
  {M.~T.}\ \bibnamefont {Lichtman}}, \bibinfo {author} {\bibfnamefont
  {Y.}~\bibnamefont {Sun}}, \bibinfo {author} {\bibfnamefont {M.}~\bibnamefont
  {Ebert}},\ and\ \bibinfo {author} {\bibfnamefont {M.}~\bibnamefont
  {Saffman}},\ }\bibfield  {title} {\bibinfo {title} {Rydberg-mediated
  entanglement in a two-dimensional neutral atom qubit array},\ }\href
  {https://doi.org/10.1103/PhysRevLett.123.230501} {\bibfield  {journal}
  {\bibinfo  {journal} {Phys. Rev. Lett.}\ }\textbf {\bibinfo {volume} {123}},\
  \bibinfo {pages} {230501} (\bibinfo {year} {2019})}\BibitemShut {NoStop}%
\bibitem [{\citenamefont {Saffman}\ \emph {et~al.}(2010)\citenamefont
  {Saffman}, \citenamefont {Walker},\ and\ \citenamefont
  {M\o{}lmer}}]{Saffman2010}%
  \BibitemOpen
  \bibfield  {author} {\bibinfo {author} {\bibfnamefont {M.}~\bibnamefont
  {Saffman}}, \bibinfo {author} {\bibfnamefont {T.~G.}\ \bibnamefont
  {Walker}},\ and\ \bibinfo {author} {\bibfnamefont {K.}~\bibnamefont
  {M\o{}lmer}},\ }\bibfield  {title} {\bibinfo {title} {Quantum information
  with {R}ydberg atoms},\ }\href {https://doi.org/10.1103/RevModPhys.82.2313}
  {\bibfield  {journal} {\bibinfo  {journal} {Rev. Mod. Phys.}\ }\textbf
  {\bibinfo {volume} {82}},\ \bibinfo {pages} {2313} (\bibinfo {year}
  {2010})}\BibitemShut {NoStop}%
\bibitem [{\citenamefont {Bluvstein}\ \emph {et~al.}(2024)\citenamefont
  {Bluvstein}, \citenamefont {Evered}, \citenamefont {Geim}, \citenamefont
  {Li}, \citenamefont {Zhou}, \citenamefont {Manovitz}, \citenamefont {Ebadi},
  \citenamefont {Cain}, \citenamefont {Kalinowski}, \citenamefont {Hangleiter},
  \citenamefont {Bonilla~Ataides}, \citenamefont {Maskara}, \citenamefont
  {Cong}, \citenamefont {Gao}, \citenamefont {Sales~Rodriguez}, \citenamefont
  {Karolyshyn}, \citenamefont {Semeghini}, \citenamefont {Gullans},
  \citenamefont {Greiner}, \citenamefont {Vuletic},\ and\ \citenamefont
  {Lukin}}]{Bluvstein2024}%
  \BibitemOpen
  \bibfield  {author} {\bibinfo {author} {\bibfnamefont {D.}~\bibnamefont
  {Bluvstein}}, \bibinfo {author} {\bibfnamefont {S.~J.}\ \bibnamefont
  {Evered}}, \bibinfo {author} {\bibfnamefont {A.~A.}\ \bibnamefont {Geim}},
  \bibinfo {author} {\bibfnamefont {S.~H.}\ \bibnamefont {Li}}, \bibinfo
  {author} {\bibfnamefont {H.}~\bibnamefont {Zhou}}, \bibinfo {author}
  {\bibfnamefont {T.}~\bibnamefont {Manovitz}}, \bibinfo {author}
  {\bibfnamefont {S.}~\bibnamefont {Ebadi}}, \bibinfo {author} {\bibfnamefont
  {M.}~\bibnamefont {Cain}}, \bibinfo {author} {\bibfnamefont {M.}~\bibnamefont
  {Kalinowski}}, \bibinfo {author} {\bibfnamefont {D.}~\bibnamefont
  {Hangleiter}}, \bibinfo {author} {\bibfnamefont {J.~P.}\ \bibnamefont
  {Bonilla~Ataides}}, \bibinfo {author} {\bibfnamefont {N.}~\bibnamefont
  {Maskara}}, \bibinfo {author} {\bibfnamefont {I.}~\bibnamefont {Cong}},
  \bibinfo {author} {\bibfnamefont {X.}~\bibnamefont {Gao}}, \bibinfo {author}
  {\bibfnamefont {P.}~\bibnamefont {Sales~Rodriguez}}, \bibinfo {author}
  {\bibfnamefont {T.}~\bibnamefont {Karolyshyn}}, \bibinfo {author}
  {\bibfnamefont {G.}~\bibnamefont {Semeghini}}, \bibinfo {author}
  {\bibfnamefont {M.~J.}\ \bibnamefont {Gullans}}, \bibinfo {author}
  {\bibfnamefont {M.}~\bibnamefont {Greiner}}, \bibinfo {author} {\bibfnamefont
  {V.}~\bibnamefont {Vuletic}},\ and\ \bibinfo {author} {\bibfnamefont {M.~D.}\
  \bibnamefont {Lukin}},\ }\bibfield  {title} {\bibinfo {title} {Logical
  quantum processor based on reconfigurable atom arrays},\ }\href
  {https://doi.org/10.1038/s41586-023-06927-3} {\bibfield  {journal} {\bibinfo
  {journal} {Nature}\ }\textbf {\bibinfo {volume} {626}},\ \bibinfo {pages}
  {58} (\bibinfo {year} {2024})}\BibitemShut {NoStop}%
\bibitem [{\citenamefont {Henriet}\ \emph {et~al.}(2020)\citenamefont
  {Henriet}, \citenamefont {Beguin}, \citenamefont {Signoles}, \citenamefont
  {Lahaye}, \citenamefont {Browaeys}, \citenamefont {Reymond},\ and\
  \citenamefont {Jurczak}}]{Henriet2020}%
  \BibitemOpen
  \bibfield  {author} {\bibinfo {author} {\bibfnamefont {L.}~\bibnamefont
  {Henriet}}, \bibinfo {author} {\bibfnamefont {L.}~\bibnamefont {Beguin}},
  \bibinfo {author} {\bibfnamefont {A.}~\bibnamefont {Signoles}}, \bibinfo
  {author} {\bibfnamefont {T.}~\bibnamefont {Lahaye}}, \bibinfo {author}
  {\bibfnamefont {A.}~\bibnamefont {Browaeys}}, \bibinfo {author}
  {\bibfnamefont {G.-O.}\ \bibnamefont {Reymond}},\ and\ \bibinfo {author}
  {\bibfnamefont {C.}~\bibnamefont {Jurczak}},\ }\bibfield  {title} {\bibinfo
  {title} {Quantum computing with neutral atoms},\ }\href
  {https://doi.org/10.22331/q-2020-09-21-327} {\bibfield  {journal} {\bibinfo
  {journal} {{Quantum}}\ }\textbf {\bibinfo {volume} {4}},\ \bibinfo {pages}
  {327} (\bibinfo {year} {2020})}\BibitemShut {NoStop}%
\bibitem [{\citenamefont {Pause}\ \emph {et~al.}(2024)\citenamefont {Pause},
  \citenamefont {Sturm}, \citenamefont {Mittenb\"{u}hler}, \citenamefont
  {Amann}, \citenamefont {Preuschoff}, \citenamefont {Sch\"{a}ffner},
  \citenamefont {Schlosser},\ and\ \citenamefont {Birkl}}]{Pause2024}%
  \BibitemOpen
  \bibfield  {author} {\bibinfo {author} {\bibfnamefont {L.}~\bibnamefont
  {Pause}}, \bibinfo {author} {\bibfnamefont {L.}~\bibnamefont {Sturm}},
  \bibinfo {author} {\bibfnamefont {M.}~\bibnamefont {Mittenb\"{u}hler}},
  \bibinfo {author} {\bibfnamefont {S.}~\bibnamefont {Amann}}, \bibinfo
  {author} {\bibfnamefont {T.}~\bibnamefont {Preuschoff}}, \bibinfo {author}
  {\bibfnamefont {D.}~\bibnamefont {Sch\"{a}ffner}}, \bibinfo {author}
  {\bibfnamefont {M.}~\bibnamefont {Schlosser}},\ and\ \bibinfo {author}
  {\bibfnamefont {G.}~\bibnamefont {Birkl}},\ }\bibfield  {title} {\bibinfo
  {title} {Supercharged two-dimensional tweezer array with more than 1000
  atomic qubits},\ }\href {https://doi.org/10.1364/OPTICA.513551} {\bibfield
  {journal} {\bibinfo  {journal} {Optica}\ }\textbf {\bibinfo {volume} {11}},\
  \bibinfo {pages} {222} (\bibinfo {year} {2024})}\BibitemShut {NoStop}%
\bibitem [{\citenamefont {Sedlacek}\ \emph {et~al.}(2012)\citenamefont
  {Sedlacek}, \citenamefont {Schwettmann}, \citenamefont {K\"{u}bler},
  \citenamefont {L\"{o}w}, \citenamefont {Pfau},\ and\ \citenamefont
  {Shaffer}}]{Sedlacek2012}%
  \BibitemOpen
  \bibfield  {author} {\bibinfo {author} {\bibfnamefont {J.~A.}\ \bibnamefont
  {Sedlacek}}, \bibinfo {author} {\bibfnamefont {A.}~\bibnamefont
  {Schwettmann}}, \bibinfo {author} {\bibfnamefont {H.}~\bibnamefont
  {K\"{u}bler}}, \bibinfo {author} {\bibfnamefont {R.}~\bibnamefont {L\"{o}w}},
  \bibinfo {author} {\bibfnamefont {T.}~\bibnamefont {Pfau}},\ and\ \bibinfo
  {author} {\bibfnamefont {J.~P.}\ \bibnamefont {Shaffer}},\ }\bibfield
  {title} {\bibinfo {title} {Microwave electrometry with {R}ydberg atoms in a
  vapour cell using bright atomic resonances},\ }\href
  {https://doi.org/10.1038/nphys2423} {\bibfield  {journal} {\bibinfo
  {journal} {Nature Phys.}\ }\textbf {\bibinfo {volume} {8}},\ \bibinfo {pages}
  {819} (\bibinfo {year} {2012})}\BibitemShut {NoStop}%
\bibitem [{\citenamefont {Fan}\ \emph {et~al.}(2015)\citenamefont {Fan},
  \citenamefont {Kumar}, \citenamefont {Sedlacek}, \citenamefont {K\"{u}bler},
  \citenamefont {Karimkashi},\ and\ \citenamefont {Shaffer}}]{Fan2015}%
  \BibitemOpen
  \bibfield  {author} {\bibinfo {author} {\bibfnamefont {H.}~\bibnamefont
  {Fan}}, \bibinfo {author} {\bibfnamefont {S.}~\bibnamefont {Kumar}}, \bibinfo
  {author} {\bibfnamefont {J.}~\bibnamefont {Sedlacek}}, \bibinfo {author}
  {\bibfnamefont {H.}~\bibnamefont {K\"{u}bler}}, \bibinfo {author}
  {\bibfnamefont {S.}~\bibnamefont {Karimkashi}},\ and\ \bibinfo {author}
  {\bibfnamefont {J.~P.}\ \bibnamefont {Shaffer}},\ }\bibfield  {title}
  {\bibinfo {title} {Atom based {RF} electric field sensing},\ }\href
  {https://doi.org/10.1088/0953-4075/48/20/202001} {\bibfield  {journal}
  {\bibinfo  {journal} {J. Phys. B: At. Mol. Opt. Phys.}\ }\textbf {\bibinfo
  {volume} {48}},\ \bibinfo {pages} {202001} (\bibinfo {year}
  {2015})}\BibitemShut {NoStop}%
\bibitem [{\citenamefont {Fancher}\ \emph {et~al.}(2021)\citenamefont
  {Fancher}, \citenamefont {Scherer}, \citenamefont {John},\ and\ \citenamefont
  {Schmittberger~Marlow}}]{Fancher2021}%
  \BibitemOpen
  \bibfield  {author} {\bibinfo {author} {\bibfnamefont {C.~T.}\ \bibnamefont
  {Fancher}}, \bibinfo {author} {\bibfnamefont {D.~R.}\ \bibnamefont
  {Scherer}}, \bibinfo {author} {\bibfnamefont {M.~C.~S.}\ \bibnamefont
  {John}},\ and\ \bibinfo {author} {\bibfnamefont {B.~L.}\ \bibnamefont
  {Schmittberger~Marlow}},\ }\bibfield  {title} {\bibinfo {title} {Rydberg atom
  electric field sensors for communications and sensing},\ }\href
  {https://doi.org/10.1109/TQE.2021.3065227} {\bibfield  {journal} {\bibinfo
  {journal} {IEEE Trans. Quantum Eng.}\ }\textbf {\bibinfo {volume} {2}},\
  \bibinfo {pages} {1} (\bibinfo {year} {2021})}\BibitemShut {NoStop}%
\bibitem [{\citenamefont {Safronova}\ \emph {et~al.}(2016)\citenamefont
  {Safronova}, \citenamefont {Safronova},\ and\ \citenamefont
  {Clark}}]{Safronova2016}%
  \BibitemOpen
  \bibfield  {author} {\bibinfo {author} {\bibfnamefont {M.~S.}\ \bibnamefont
  {Safronova}}, \bibinfo {author} {\bibfnamefont {U.~I.}\ \bibnamefont
  {Safronova}},\ and\ \bibinfo {author} {\bibfnamefont {C.~W.}\ \bibnamefont
  {Clark}},\ }\bibfield  {title} {\bibinfo {title} {Magic wavelengths, matrix
  elements, polarizabilities, and lifetimes of {C}s},\ }\href
  {https://doi.org/10.1103/PhysRevA.94.012505} {\bibfield  {journal} {\bibinfo
  {journal} {Phys. Rev. A}\ }\textbf {\bibinfo {volume} {94}},\ \bibinfo
  {pages} {012505} (\bibinfo {year} {2016})}\BibitemShut {NoStop}%
\bibitem [{\citenamefont {Dalibard}\ and\ \citenamefont
  {Cohen-Tannoudji}(1989)}]{Dalibard1989}%
  \BibitemOpen
  \bibfield  {author} {\bibinfo {author} {\bibfnamefont {J.}~\bibnamefont
  {Dalibard}}\ and\ \bibinfo {author} {\bibfnamefont {C.}~\bibnamefont
  {Cohen-Tannoudji}},\ }\bibfield  {title} {\bibinfo {title} {Laser cooling
  below the {D}oppler limit by polarization gradients: {S}imple theoretical
  models},\ }\href {https://doi.org/10.1364/JOSAB.6.002023} {\bibfield
  {journal} {\bibinfo  {journal} {J. Opt. Soc. Am. B}\ }\textbf {\bibinfo
  {volume} {6}},\ \bibinfo {pages} {2023} (\bibinfo {year} {1989})}\BibitemShut
  {NoStop}%
\bibitem [{\citenamefont {Zhong-Hua}\ \emph {et~al.}(2014)\citenamefont
  {Zhong-Hua}, \citenamefont {Jin-Peng}, \citenamefont {Yan-Ting},
  \citenamefont {Xue-Fang}, \citenamefont {Lian-Tuan},\ and\ \citenamefont
  {Suo-Tang}}]{Ji2014}%
  \BibitemOpen
  \bibfield  {author} {\bibinfo {author} {\bibfnamefont {J.}~\bibnamefont
  {Zhong-Hua}}, \bibinfo {author} {\bibfnamefont {Y.}~\bibnamefont {Jin-Peng}},
  \bibinfo {author} {\bibfnamefont {Z.}~\bibnamefont {Yan-Ting}}, \bibinfo
  {author} {\bibfnamefont {C.}~\bibnamefont {Xue-Fang}}, \bibinfo {author}
  {\bibfnamefont {X.}~\bibnamefont {Lian-Tuan}},\ and\ \bibinfo {author}
  {\bibfnamefont {J.}~\bibnamefont {Suo-Tang}},\ }\bibfield  {title} {\bibinfo
  {title} {Systematically investigating the polarization gradient cooling in an
  optical molasses of ultracold cesium atoms},\ }\href
  {https://doi.org/10.1088/1674-1056/23/11/113702} {\bibfield  {journal}
  {\bibinfo  {journal} {Chin. Phys. B}\ }\textbf {\bibinfo {volume} {23}},\
  \bibinfo {pages} {113702} (\bibinfo {year} {2014})}\BibitemShut {NoStop}%
\bibitem [{SM()}]{SM}%
  \BibitemOpen
  \href@noop {} {}\bibinfo {howpublished} {See Supplemental
  Material at [url], which includes Refs. \cite{Bevan_1912,Kratz_1949,McNally_1949,
Kleiman_1962,Bockasten_1964,Eriksson_1970,
Lorenzen_1979,Lorenzen_1980,Sansonetti_1981,OSullivan_1983,Gerginov2003,
Rahaman2024,Dedman2001,Fabrikant1986,Aymar_1996} for additional information about the spctrum fitting methods, uncertainty analysis and a detailed discussion of the theoretical calculations of electric dipole matrix elements.}\BibitemShut {NoStop}%
\bibitem [{\citenamefont {Bevan}(1912)}]{Bevan_1912}%
  \BibitemOpen
  \bibfield  {author} {\bibinfo {author} {\bibfnamefont {P.~V.}\ \bibnamefont
  {Bevan}},\ }\bibfield  {title} {\bibinfo {title} {Spetroscopic observations:
  Lithium and c\oe{}sium},\ }\href {https://doi.org/10.1098/rspa.1912.0023}
  {\bibfield  {journal} {\bibinfo  {journal} {Proc. R. Soc. A}\ }\textbf
  {\bibinfo {volume} {86}},\ \bibinfo {pages} {320} (\bibinfo {year}
  {1912})}\BibitemShut {NoStop}%
\bibitem [{\citenamefont {Kratz}(1949)}]{Kratz_1949}%
  \BibitemOpen
  \bibfield  {author} {\bibinfo {author} {\bibfnamefont {H.~R.}\ \bibnamefont
  {Kratz}},\ }\bibfield  {title} {\bibinfo {title} {The principal series of
  potassium, rubidium, and cesium in absorption},\ }\href
  {https://doi.org/10.1103/PhysRev.75.1844} {\bibfield  {journal} {\bibinfo
  {journal} {Phys. Rev.}\ }\textbf {\bibinfo {volume} {75}},\ \bibinfo {pages}
  {1844} (\bibinfo {year} {1949})}\BibitemShut {NoStop}%
\bibitem [{\citenamefont {McNally}\ \emph {et~al.}(1949)\citenamefont
  {McNally}, \citenamefont {Molnar}, \citenamefont {Hitchcock},\ and\
  \citenamefont {Oliver}}]{McNally_1949}%
  \BibitemOpen
  \bibfield  {author} {\bibinfo {author} {\bibfnamefont {J.~R.}\ \bibnamefont
  {McNally}}, \bibinfo {author} {\bibfnamefont {J.~P.}\ \bibnamefont {Molnar}},
  \bibinfo {author} {\bibfnamefont {W.~J.}\ \bibnamefont {Hitchcock}},\ and\
  \bibinfo {author} {\bibfnamefont {N.~F.}\ \bibnamefont {Oliver}},\ }\bibfield
   {title} {\bibinfo {title} {High members of the principal series in
  caesium},\ }\href {https://doi.org/10.1364/JOSA.39.000057} {\bibfield
  {journal} {\bibinfo  {journal} {J. Opt. Soc. Am.}\ }\textbf {\bibinfo
  {volume} {39}},\ \bibinfo {pages} {57} (\bibinfo {year} {1949})}\BibitemShut
  {NoStop}%
\bibitem [{\citenamefont {Kleiman}(1962)}]{Kleiman_1962}%
  \BibitemOpen
  \bibfield  {author} {\bibinfo {author} {\bibfnamefont {H.}~\bibnamefont
  {Kleiman}},\ }\bibfield  {title} {\bibinfo {title} {Interferometric
  measurements of cesium {I}},\ }\href {https://doi.org/10.1364/JOSA.52.000441}
  {\bibfield  {journal} {\bibinfo  {journal} {J. Opt. Soc. Am.}\ }\textbf
  {\bibinfo {volume} {52}},\ \bibinfo {pages} {441} (\bibinfo {year}
  {1962})}\BibitemShut {NoStop}%
\bibitem [{\citenamefont {Bockasten}(1964)}]{Bockasten_1964}%
  \BibitemOpen
  \bibfield  {author} {\bibinfo {author} {\bibfnamefont {K.}~\bibnamefont
  {Bockasten}},\ }\bibfield  {title} {\bibinfo {title} {Polarization formula
  for the $^2{F}$-series of {C}s{I}},\ }\href
  {https://doi.org/10.1364/JOSA.54.001065} {\bibfield  {journal} {\bibinfo
  {journal} {J. Opt. Soc. Am.}\ }\textbf {\bibinfo {volume} {54}},\ \bibinfo
  {pages} {1065} (\bibinfo {year} {1964})}\BibitemShut {NoStop}%
\bibitem [{\citenamefont {Eriksson}\ and\ \citenamefont
  {Wen{\aa}ker}(1970)}]{Eriksson_1970}%
  \BibitemOpen
  \bibfield  {author} {\bibinfo {author} {\bibfnamefont {K.~B.~S.}\
  \bibnamefont {Eriksson}}\ and\ \bibinfo {author} {\bibfnamefont
  {I.}~\bibnamefont {Wen{\aa}ker}},\ }\bibfield  {title} {\bibinfo {title} {New
  wavelength measurements in {C}s {I}},\ }\href
  {https://doi.org/10.1088/0031-8949/1/1/003} {\bibfield  {journal} {\bibinfo
  {journal} {Phys. Scr.}\ }\textbf {\bibinfo {volume} {1}},\ \bibinfo {pages}
  {21} (\bibinfo {year} {1970})}\BibitemShut {NoStop}%
\bibitem [{\citenamefont {Lorenzen}\ and\ \citenamefont
  {Niemax}(1979)}]{Lorenzen_1979}%
  \BibitemOpen
  \bibfield  {author} {\bibinfo {author} {\bibfnamefont {C.-J.}\ \bibnamefont
  {Lorenzen}}\ and\ \bibinfo {author} {\bibfnamefont {K.}~\bibnamefont
  {Niemax}},\ }\bibfield  {title} {\bibinfo {title} {New measurements of the
  $n^2{P}_{1/2,3/2}$ energy levels of {C}s({I})},\ }\href
  {https://doi.org/https://doi.org/10.1016/0022-4073(79)90115-8} {\bibfield
  {journal} {\bibinfo  {journal} {J. Quant. Spectrosc. Radiat. Transf.}\
  }\textbf {\bibinfo {volume} {22}},\ \bibinfo {pages} {247} (\bibinfo {year}
  {1979})}\BibitemShut {NoStop}%
\bibitem [{\citenamefont {Lorenzen}\ \emph {et~al.}(1980)\citenamefont
  {Lorenzen}, \citenamefont {Weber},\ and\ \citenamefont
  {Niemax}}]{Lorenzen_1980}%
  \BibitemOpen
  \bibfield  {author} {\bibinfo {author} {\bibfnamefont {C.-J.}\ \bibnamefont
  {Lorenzen}}, \bibinfo {author} {\bibfnamefont {K.-H.}\ \bibnamefont
  {Weber}},\ and\ \bibinfo {author} {\bibfnamefont {K.}~\bibnamefont
  {Niemax}},\ }\bibfield  {title} {\bibinfo {title} {Energies of the
  $n^2{S}_{1/2}$ and $n^2{D}_{3/2,5/2}$ states of {C}s},\ }\href
  {https://doi.org/https://doi.org/10.1016/0030-4018(80)90242-4} {\bibfield
  {journal} {\bibinfo  {journal} {Opt. Commun.}\ }\textbf {\bibinfo {volume}
  {33}},\ \bibinfo {pages} {271} (\bibinfo {year} {1980})}\BibitemShut
  {NoStop}%
\bibitem [{\citenamefont {Sansonetti}\ \emph {et~al.}(1981)\citenamefont
  {Sansonetti}, \citenamefont {Andrew},\ and\ \citenamefont
  {Verges}}]{Sansonetti_1981}%
  \BibitemOpen
  \bibfield  {author} {\bibinfo {author} {\bibfnamefont {C.~J.}\ \bibnamefont
  {Sansonetti}}, \bibinfo {author} {\bibfnamefont {K.~L.}\ \bibnamefont
  {Andrew}},\ and\ \bibinfo {author} {\bibfnamefont {J.}~\bibnamefont
  {Verges}},\ }\bibfield  {title} {\bibinfo {title} {Polarization, penetration,
  and exchange effects in the hydrogenlike $nf$ and $ng$ terms of cesium},\
  }\href {https://doi.org/10.1364/JOSA.71.000423} {\bibfield  {journal}
  {\bibinfo  {journal} {J. Opt. Soc. Am.}\ }\textbf {\bibinfo {volume} {71}},\
  \bibinfo {pages} {423} (\bibinfo {year} {1981})}\BibitemShut {NoStop}%
\bibitem [{\citenamefont {O'Sullivan}\ and\ \citenamefont
  {Stoicheff}(1983)}]{OSullivan_1983}%
  \BibitemOpen
  \bibfield  {author} {\bibinfo {author} {\bibfnamefont {M.~S.}\ \bibnamefont
  {O'Sullivan}}\ and\ \bibinfo {author} {\bibfnamefont {B.~P.}\ \bibnamefont
  {Stoicheff}},\ }\bibfield  {title} {\bibinfo {title} {Doppler-free two-photon
  absorption spectrum of cesium},\ }\href {https://doi.org/10.1139/p83-116}
  {\bibfield  {journal} {\bibinfo  {journal} {Can. J. Phys.}\ }\textbf
  {\bibinfo {volume} {61}},\ \bibinfo {pages} {940} (\bibinfo {year}
  {1983})}\BibitemShut {NoStop}%
\bibitem [{\citenamefont {Gerginov}\ \emph {et~al.}(2003)\citenamefont
  {Gerginov}, \citenamefont {Derevianko},\ and\ \citenamefont
  {Tanner}}]{Gerginov2003}%
  \BibitemOpen
  \bibfield  {author} {\bibinfo {author} {\bibfnamefont {V.}~\bibnamefont
  {Gerginov}}, \bibinfo {author} {\bibfnamefont {A.}~\bibnamefont
  {Derevianko}},\ and\ \bibinfo {author} {\bibfnamefont {C.~E.}\ \bibnamefont
  {Tanner}},\ }\bibfield  {title} {\bibinfo {title} {Observation of the nuclear
  magnetic octupole moment of $^{133}\mathrm{C}\mathrm{s}$},\ }\href
  {https://doi.org/10.1103/PhysRevLett.91.072501} {\bibfield  {journal}
  {\bibinfo  {journal} {Phys. Rev. Lett.}\ }\textbf {\bibinfo {volume} {91}},\
  \bibinfo {pages} {072501} (\bibinfo {year} {2003})}\BibitemShut {NoStop}%
\bibitem [{\citenamefont {Rahaman}\ \emph {et~al.}(2024)\citenamefont
  {Rahaman}, \citenamefont {Wright},\ and\ \citenamefont
  {Dutta}}]{Rahaman2024}%
  \BibitemOpen
  \bibfield  {author} {\bibinfo {author} {\bibfnamefont {B.}~\bibnamefont
  {Rahaman}}, \bibinfo {author} {\bibfnamefont {S.~C.}\ \bibnamefont
  {Wright}},\ and\ \bibinfo {author} {\bibfnamefont {S.}~\bibnamefont
  {Dutta}},\ }\bibfield  {title} {\bibinfo {title} {Observation of quantum
  interference of optical transition pathways in {D}oppler-free two-photon
  spectroscopy and implications for precision measurements},\ }\href
  {https://doi.org/10.1103/PhysRevA.109.042820} {\bibfield  {journal} {\bibinfo
   {journal} {Phys. Rev. A}\ }\textbf {\bibinfo {volume} {109}},\ \bibinfo
  {pages} {042820} (\bibinfo {year} {2024})}\BibitemShut {NoStop}%
\bibitem [{\citenamefont {Dedman}\ \emph {et~al.}(2001)\citenamefont {Dedman},
  \citenamefont {Baldwin},\ and\ \citenamefont {Colla}}]{Dedman2001}%
  \BibitemOpen
  \bibfield  {author} {\bibinfo {author} {\bibfnamefont {C.~J.}\ \bibnamefont
  {Dedman}}, \bibinfo {author} {\bibfnamefont {K.~G.~H.}\ \bibnamefont
  {Baldwin}},\ and\ \bibinfo {author} {\bibfnamefont {M.}~\bibnamefont
  {Colla}},\ }\bibfield  {title} {\bibinfo {title} {{Fast switching of magnetic
  fields in a magneto-optic trap}},\ }\href {https://doi.org/10.1063/1.1408935}
  {\bibfield  {journal} {\bibinfo  {journal} {Review of Scientific
  Instruments}\ }\textbf {\bibinfo {volume} {72}},\ \bibinfo {pages} {4055}
  (\bibinfo {year} {2001})}\BibitemShut {NoStop}%
\bibitem [{\citenamefont {Fabrikant}(1986)}]{Fabrikant1986}%
  \BibitemOpen
  \bibfield  {author} {\bibinfo {author} {\bibfnamefont {I.~I.}\ \bibnamefont
  {Fabrikant}},\ }\bibfield  {title} {\bibinfo {title} {Interaction of
  {R}ydberg atoms and thermal electrons with {K}, {R}b and {C}s atoms},\ }\href
  {https://doi.org/10.1088/0022-3700/19/10/021} {\bibfield  {journal} {\bibinfo
   {journal} {J. Phys. B: At. Mol. Opt. Phys.}\ }\textbf {\bibinfo {volume}
  {19}},\ \bibinfo {pages} {1527} (\bibinfo {year} {1986})}\BibitemShut
  {NoStop}%
\bibitem [{\citenamefont {Aymar}\ \emph {et~al.}(1996)\citenamefont {Aymar},
  \citenamefont {Greene},\ and\ \citenamefont {Luc-Koenig}}]{Aymar_1996}%
  \BibitemOpen
  \bibfield  {author} {\bibinfo {author} {\bibfnamefont {M.}~\bibnamefont
  {Aymar}}, \bibinfo {author} {\bibfnamefont {C.~H.}\ \bibnamefont {Greene}},\
  and\ \bibinfo {author} {\bibfnamefont {E.}~\bibnamefont {Luc-Koenig}},\
  }\bibfield  {title} {\bibinfo {title} {Multichannel {R}ydberg spectroscopy of
  complex atoms},\ }\href {https://doi.org/10.1103/RevModPhys.68.1015}
  {\bibfield  {journal} {\bibinfo  {journal} {Rev. Mod. Phys.}\ }\textbf
  {\bibinfo {volume} {68}},\ \bibinfo {pages} {1015} (\bibinfo {year}
  {1996})}\BibitemShut {NoStop}%
\bibitem [{\citenamefont {\v{S}ibali\'c}\ \emph {et~al.}(2017)\citenamefont
  {\v{S}ibali\'c}, \citenamefont {Pritchard}, \citenamefont {Adams},\ and\
  \citenamefont {Weatherill}}]{ARC2017}%
  \BibitemOpen
  \bibfield  {author} {\bibinfo {author} {\bibfnamefont {N.}~\bibnamefont
  {\v{S}ibali\'c}}, \bibinfo {author} {\bibfnamefont {J.~D.}\ \bibnamefont
  {Pritchard}}, \bibinfo {author} {\bibfnamefont {C.~S.}\ \bibnamefont
  {Adams}},\ and\ \bibinfo {author} {\bibfnamefont {K.~J.}\ \bibnamefont
  {Weatherill}},\ }\bibfield  {title} {\bibinfo {title} {{ARC}: An open-source
  library for calculating properties of alkali {R}ydberg atoms},\ }\href
  {https://doi.org/https://doi.org/10.1016/j.cpc.2017.06.015} {\bibfield
  {journal} {\bibinfo  {journal} {Comput. Phys. Commun.}\ }\textbf {\bibinfo
  {volume} {220}},\ \bibinfo {pages} {319} (\bibinfo {year}
  {2017})}\BibitemShut {NoStop}%
\bibitem [{\citenamefont {Meyer}\ \emph {et~al.}(2020)\citenamefont {Meyer},
  \citenamefont {Castillo}, \citenamefont {Cox},\ and\ \citenamefont
  {Kunz}}]{Meyer2020}%
  \BibitemOpen
  \bibfield  {author} {\bibinfo {author} {\bibfnamefont {D.~H.}\ \bibnamefont
  {Meyer}}, \bibinfo {author} {\bibfnamefont {Z.~A.}\ \bibnamefont {Castillo}},
  \bibinfo {author} {\bibfnamefont {K.~C.}\ \bibnamefont {Cox}},\ and\ \bibinfo
  {author} {\bibfnamefont {P.~D.}\ \bibnamefont {Kunz}},\ }\bibfield  {title}
  {\bibinfo {title} {Assessment of {R}ydberg atoms for wideband electric field
  sensing},\ }\href {https://doi.org/10.1088/1361-6455/ab6051} {\bibfield
  {journal} {\bibinfo  {journal} {J. Phys. B: At. Mol. Opt. Phys.}\ }\textbf
  {\bibinfo {volume} {53}},\ \bibinfo {pages} {034001} (\bibinfo {year}
  {2020})}\BibitemShut {NoStop}%
\bibitem [{\citenamefont {Bai}\ \emph {et~al.}(2020)\citenamefont {Bai},
  \citenamefont {Bai}, \citenamefont {Han}, \citenamefont {Jiao}, \citenamefont
  {Zhao},\ and\ \citenamefont {Jia}}]{Bai2020}%
  \BibitemOpen
  \bibfield  {author} {\bibinfo {author} {\bibfnamefont {J.}~\bibnamefont
  {Bai}}, \bibinfo {author} {\bibfnamefont {S.}~\bibnamefont {Bai}}, \bibinfo
  {author} {\bibfnamefont {X.}~\bibnamefont {Han}}, \bibinfo {author}
  {\bibfnamefont {Y.}~\bibnamefont {Jiao}}, \bibinfo {author} {\bibfnamefont
  {J.}~\bibnamefont {Zhao}},\ and\ \bibinfo {author} {\bibfnamefont
  {S.}~\bibnamefont {Jia}},\ }\bibfield  {title} {\bibinfo {title} {Precise
  measurements of polarizabilities of cesium $n{S}$ {R}ydberg states in an
  ultra-cold atomic ensemble},\ }\href
  {https://doi.org/10.1088/1367-2630/abaf30} {\bibfield  {journal} {\bibinfo
  {journal} {New J. Phys.}\ }\textbf {\bibinfo {volume} {22}},\ \bibinfo
  {pages} {093032} (\bibinfo {year} {2020})}\BibitemShut {NoStop}%
\bibitem [{\citenamefont {Auzinsh}\ \emph {et~al.}(2007)\citenamefont
  {Auzinsh}, \citenamefont {Bluss}, \citenamefont {Ferber}, \citenamefont
  {Gahbauer}, \citenamefont {Jarmola}, \citenamefont {Safronova}, \citenamefont
  {Safronova},\ and\ \citenamefont {Tamanis}}]{Auzinsh2007}%
  \BibitemOpen
  \bibfield  {author} {\bibinfo {author} {\bibfnamefont {M.}~\bibnamefont
  {Auzinsh}}, \bibinfo {author} {\bibfnamefont {K.}~\bibnamefont {Bluss}},
  \bibinfo {author} {\bibfnamefont {R.}~\bibnamefont {Ferber}}, \bibinfo
  {author} {\bibfnamefont {F.}~\bibnamefont {Gahbauer}}, \bibinfo {author}
  {\bibfnamefont {A.}~\bibnamefont {Jarmola}}, \bibinfo {author} {\bibfnamefont
  {M.~S.}\ \bibnamefont {Safronova}}, \bibinfo {author} {\bibfnamefont {U.~I.}\
  \bibnamefont {Safronova}},\ and\ \bibinfo {author} {\bibfnamefont
  {M.}~\bibnamefont {Tamanis}},\ }\bibfield  {title} {\bibinfo {title}
  {Level-crossing spectroscopy of the 7, 9, and $10{D}_{5/2}$ states of
  $^{133}\mathrm{Cs}$ and validation of relativistic many-body calculations of
  the polarizabilities and hyperfine constants},\ }\href
  {https://doi.org/10.1103/PhysRevA.75.022502} {\bibfield  {journal} {\bibinfo
  {journal} {Phys. Rev. A}\ }\textbf {\bibinfo {volume} {75}},\ \bibinfo
  {pages} {022502} (\bibinfo {year} {2007})}\BibitemShut {NoStop}%
\bibitem [{\citenamefont {Sa\ss{}mannshausen}\ \emph
  {et~al.}(2013)\citenamefont {Sa\ss{}mannshausen}, \citenamefont {Merkt},\
  and\ \citenamefont {Deiglmayr}}]{Sassmann2013}%
  \BibitemOpen
  \bibfield  {author} {\bibinfo {author} {\bibfnamefont {H.}~\bibnamefont
  {Sa\ss{}mannshausen}}, \bibinfo {author} {\bibfnamefont {F.}~\bibnamefont
  {Merkt}},\ and\ \bibinfo {author} {\bibfnamefont {J.}~\bibnamefont
  {Deiglmayr}},\ }\bibfield  {title} {\bibinfo {title} {High-resolution
  spectroscopy of {R}ydberg states in an ultracold cesium gas},\ }\href
  {https://doi.org/10.1103/PhysRevA.87.032519} {\bibfield  {journal} {\bibinfo
  {journal} {Phys. Rev. A}\ }\textbf {\bibinfo {volume} {87}},\ \bibinfo
  {pages} {032519} (\bibinfo {year} {2013})}\BibitemShut {NoStop}%
\bibitem [{\citenamefont {Weber}\ and\ \citenamefont
  {Sansonetti}(1987)}]{Weber1987}%
  \BibitemOpen
  \bibfield  {author} {\bibinfo {author} {\bibfnamefont {K.-H.}\ \bibnamefont
  {Weber}}\ and\ \bibinfo {author} {\bibfnamefont {C.~J.}\ \bibnamefont
  {Sansonetti}},\ }\bibfield  {title} {\bibinfo {title} {Accurate energies of
  $n{S}$, $n{P}$, $n{D}$, $n{F}$, and $n{G}$ levels of neutral cesium},\ }\href
  {https://doi.org/10.1103/PhysRevA.35.4650} {\bibfield  {journal} {\bibinfo
  {journal} {Phys. Rev. A}\ }\textbf {\bibinfo {volume} {35}},\ \bibinfo
  {pages} {4650} (\bibinfo {year} {1987})}\BibitemShut {NoStop}%
\bibitem [{\citenamefont {Lorenzen}\ and\ \citenamefont
  {Niemax}(1984)}]{Lorenzen1984}%
  \BibitemOpen
  \bibfield  {author} {\bibinfo {author} {\bibfnamefont {C.-J.}\ \bibnamefont
  {Lorenzen}}\ and\ \bibinfo {author} {\bibfnamefont {K.}~\bibnamefont
  {Niemax}},\ }\bibfield  {title} {\bibinfo {title} {Precise quantum defects of
  $n{S}$, $n{P}$ and $n{D}$ levels in {C}s {I}},\ }\href
  {https://doi.org/10.1007/BF01419370} {\bibfield  {journal} {\bibinfo
  {journal} {Z. Phys. A}\ }\textbf {\bibinfo {volume} {315}},\ \bibinfo {pages}
  {127} (\bibinfo {year} {1984})}\BibitemShut {NoStop}%
\bibitem [{CIA()}]{CIAAW}%
  \BibitemOpen
  \href@noop {} {}\bibinfo {howpublished} {Standard atomic weights of the
  elements 2020, available online at \href{http://ciaaw.org/atomic-weights
  }{\tt{http://ciaaw.org/atomic-weights}}.}\BibitemShut {Stop}%
\bibitem [{\citenamefont {Steck}()}]{steckCsdata}%
  \BibitemOpen
  \bibfield  {author} {\bibinfo {author} {\bibfnamefont {D.}~\bibnamefont
  {Steck}},\ }\href@noop {} {}\bibinfo {howpublished} {Cesium D line data,
  available online at
  \href{http://steck.us/alkalidata}{\tt{http://steck.us/alkalidata}} (revision
  2.0.1, 2 May 2008)}\BibitemShut {NoStop}%
\bibitem [{\citenamefont {Deiglmayr}\ \emph {et~al.}(2016)\citenamefont
  {Deiglmayr}, \citenamefont {Herburger}, \citenamefont {Sa\ss{}mannshausen},
  \citenamefont {Jansen}, \citenamefont {Schmutz},\ and\ \citenamefont
  {Merkt}}]{Deiglmayr2016}%
  \BibitemOpen
  \bibfield  {author} {\bibinfo {author} {\bibfnamefont {J.}~\bibnamefont
  {Deiglmayr}}, \bibinfo {author} {\bibfnamefont {H.}~\bibnamefont
  {Herburger}}, \bibinfo {author} {\bibfnamefont {H.}~\bibnamefont
  {Sa\ss{}mannshausen}}, \bibinfo {author} {\bibfnamefont {P.}~\bibnamefont
  {Jansen}}, \bibinfo {author} {\bibfnamefont {H.}~\bibnamefont {Schmutz}},\
  and\ \bibinfo {author} {\bibfnamefont {F.}~\bibnamefont {Merkt}},\ }\bibfield
   {title} {\bibinfo {title} {Precision measurement of the ionization energy of
  {C}s {I}},\ }\href {https://doi.org/10.1103/PhysRevA.93.013424} {\bibfield
  {journal} {\bibinfo  {journal} {Phys. Rev. A}\ }\textbf {\bibinfo {volume}
  {93}},\ \bibinfo {pages} {013424} (\bibinfo {year} {2016})}\BibitemShut
  {NoStop}%
\bibitem [{\citenamefont {Lorenzen}\ and\ \citenamefont
  {Niemax}(1983)}]{Lorenzen1983}%
  \BibitemOpen
  \bibfield  {author} {\bibinfo {author} {\bibfnamefont {C.-J.}\ \bibnamefont
  {Lorenzen}}\ and\ \bibinfo {author} {\bibfnamefont {K.}~\bibnamefont
  {Niemax}},\ }\bibfield  {title} {\bibinfo {title} {Non-monotonic variation of
  the quantum defect in {C}s $n{D}_{J}$ term series},\ }\href
  {https://doi.org/10.1007/BF01415112} {\bibfield  {journal} {\bibinfo
  {journal} {Z. Phys. A}\ }\textbf {\bibinfo {volume} {311}},\ \bibinfo {pages}
  {249} (\bibinfo {year} {1983})}\BibitemShut {NoStop}%
\bibitem [{\citenamefont {Marinescu}\ \emph {et~al.}(1994)\citenamefont
  {Marinescu}, \citenamefont {Sadeghpour},\ and\ \citenamefont
  {Dalgarno}}]{Marinescu1994}%
  \BibitemOpen
  \bibfield  {author} {\bibinfo {author} {\bibfnamefont {M.}~\bibnamefont
  {Marinescu}}, \bibinfo {author} {\bibfnamefont {H.~R.}\ \bibnamefont
  {Sadeghpour}},\ and\ \bibinfo {author} {\bibfnamefont {A.}~\bibnamefont
  {Dalgarno}},\ }\bibfield  {title} {\bibinfo {title} {Dispersion coefficients
  for alkali-metal dimers},\ }\href {https://doi.org/10.1103/PhysRevA.49.982}
  {\bibfield  {journal} {\bibinfo  {journal} {Phys. Rev. A}\ }\textbf {\bibinfo
  {volume} {49}},\ \bibinfo {pages} {982} (\bibinfo {year} {1994})}\BibitemShut
  {NoStop}%
\bibitem [{\citenamefont {Drake}\ and\ \citenamefont
  {Swainson}(1991)}]{Drake1991}%
  \BibitemOpen
  \bibfield  {author} {\bibinfo {author} {\bibfnamefont {G.~W.~F.}\
  \bibnamefont {Drake}}\ and\ \bibinfo {author} {\bibfnamefont {R.~A.}\
  \bibnamefont {Swainson}},\ }\bibfield  {title} {\bibinfo {title} {Quantum
  defects and the $1/n$ dependence of {R}ydberg energies: Second-order
  polarization effects},\ }\href {https://doi.org/10.1103/PhysRevA.44.5448}
  {\bibfield  {journal} {\bibinfo  {journal} {Phys. Rev. A}\ }\textbf {\bibinfo
  {volume} {44}},\ \bibinfo {pages} {5448} (\bibinfo {year}
  {1991})}\BibitemShut {NoStop}%
\bibitem [{\citenamefont {Goy}\ \emph {et~al.}(1982)\citenamefont {Goy},
  \citenamefont {Raimond}, \citenamefont {Vitrant},\ and\ \citenamefont
  {Haroche}}]{Goy_PRA_1982}%
  \BibitemOpen
  \bibfield  {author} {\bibinfo {author} {\bibfnamefont {P.}~\bibnamefont
  {Goy}}, \bibinfo {author} {\bibfnamefont {J.~M.}\ \bibnamefont {Raimond}},
  \bibinfo {author} {\bibfnamefont {G.}~\bibnamefont {Vitrant}},\ and\ \bibinfo
  {author} {\bibfnamefont {S.}~\bibnamefont {Haroche}},\ }\bibfield  {title}
  {\bibinfo {title} {Millimeter-wave spectroscopy in cesium {R}ydberg states.
  {Q}uantum defects, fine- and hyperfine-structure measurements},\ }\href
  {https://doi.org/10.1103/PhysRevA.26.2733} {\bibfield  {journal} {\bibinfo
  {journal} {Phys. Rev. A}\ }\textbf {\bibinfo {volume} {26}},\ \bibinfo
  {pages} {2733} (\bibinfo {year} {1982})}\BibitemShut {NoStop}%
\bibitem [{\citenamefont {Simons}\ \emph {et~al.}(2016)\citenamefont {Simons},
  \citenamefont {Gordon},\ and\ \citenamefont {Holloway}}]{Simons2016}%
  \BibitemOpen
  \bibfield  {author} {\bibinfo {author} {\bibfnamefont {M.~T.}\ \bibnamefont
  {Simons}}, \bibinfo {author} {\bibfnamefont {J.~A.}\ \bibnamefont {Gordon}},\
  and\ \bibinfo {author} {\bibfnamefont {C.~L.}\ \bibnamefont {Holloway}},\
  }\bibfield  {title} {\bibinfo {title} {{Simultaneous use of Cs and Rb Rydberg
  atoms for dipole moment assessment and RF electric field measurements via
  electromagnetically induced transparency}},\ }\href
  {https://doi.org/10.1063/1.4963106} {\bibfield  {journal} {\bibinfo
  {journal} {J. Appl. Phys.}\ }\textbf {\bibinfo {volume} {120}},\ \bibinfo
  {pages} {123103} (\bibinfo {year} {2016})}\BibitemShut {NoStop}%
\end{thebibliography}
\end{document}